\documentclass[sigplan,screen]{acmart}

\usepackage{amsmath,amsfonts}
\usepackage{algorithmic}
\usepackage{textcomp}
\usepackage{xcolor}
\def\BibTeX{{\rm B\kern-.05em{\sc i\kern-.025em b}\kern-.08em
    T\kern-.1667em\lower.7ex\hbox{E}\kern-.125emX}}

\usepackage[final]{pdfpages}
\usepackage{hyperref}
\usepackage{algorithmic}
\usepackage{graphicx}
\usepackage{textcomp}
\usepackage{xcolor}
\usepackage{color}
\usepackage{url}
\usepackage{enumitem}
\usepackage{multirow}
\usepackage[ruled, vlined, linesnumbered]{algorithm2e}
\usepackage{subcaption}
\SetAlCapNameFnt{\scriptsize}
\SetAlCapFnt{\small}

\usepackage{listings}
\usepackage{verbatim}

\usepackage{xspace}

\usepackage{soul}

\newcommand{\tool}{\textsc{Once4All}}

\newcommand{\reported}{45}
\newcommand{\confirmed}{43}
\newcommand{\fixed}{40}

\newcommand*{\eg}{e.g., }
\newcommand*{\ie}{i.e., }

\usepackage{stackengine}
\usepackage{booktabs}

\newcommand{\mypara}[1]{\vspace{3pt}\noindent\textit{\textbf{#1} }}

\usepackage[most]{tcolorbox}
\usepackage{colortbl}

\usepackage{balance}
\usepackage{colortbl}
\usepackage{pgfplots}
\usepackage{microtype}
\usepgfplotslibrary{dateplot}
\pgfplotsset{compat=newest}
\usetikzlibrary{patterns}
\colorlet{shadecolor}{gray!40}
\usepackage{tikz}
\usetikzlibrary{shapes.geometric,arrows,shapes,shadows} 

\usepackage{pifont}
\usepackage{flushend}
\usepackage{ulem}
\definecolor{promptbg}{HTML}{FFF4E6}
\newtcolorbox{promptFrame}{
  colback=green!5!white,
  colframe=green!40!black,
  fonttitle=\bfseries,
  coltitle=white,
  title=Prompt,
  boxrule=1pt,
  arc=3mm,
  top=3mm, bottom=3mm,
  left=4mm, right=4mm,
  enhanced,
  attach boxed title to top left={xshift=6mm,yshift=-3mm},
  boxed title style={size=small,colback=green!60!black,arc=2mm},
  drop shadow={opacity=0.3,shadow xshift=1mm,shadow yshift=-1mm}
}

\usepackage{fbox}
\newcounter{observation}
\newtcolorbox[use counter=observation]{observationbox}{
  colback=blue!5!white,       %
  colframe=blue!40!black,     %
  fonttitle=\bfseries,        %
  coltitle=white,             %
  title=Observation~\theobservation, %
  boxrule=1pt,                %
  arc=3mm,                    %
  top=3mm, bottom=3mm,        %
  left=4mm, right=4mm,        %
  enhanced,
  attach boxed title to top left={xshift=6mm,yshift=-3mm},
  boxed title style={size=small,colback=blue!60!black,arc=2mm},
  drop shadow={opacity=0.3,shadow xshift=1mm,shadow yshift=-1mm}
}

\newcounter{finding}
\newtcolorbox[use counter=finding]{findingbox}{
  colback=green!5!white,      %
  colframe=green!40!black,    %
  fonttitle=\bfseries,        %
  coltitle=white,             %
  title=Finding~\thefinding,  %
  boxrule=1pt,                %
  arc=3mm,                    %
  top=3mm, bottom=3mm,        %
  left=4mm, right=4mm,        %
  enhanced,
  attach boxed title to top left={xshift=6mm,yshift=-3mm},
  boxed title style={size=small,colback=green!60!black,arc=2mm},
  drop shadow={opacity=0.3,shadow xshift=1mm,shadow yshift=-1mm}
}

\definecolor{mygreen}{HTML}{02818a}
\definecolor{mypurple}{HTML}{8f3c8c}

\newcommand{\keepnotes}{true}

\ifthenelse{\isundefined{\keepnotes}}{
\newcommand{\mytodogreen}[1]{}
\newcommand{\civi}[1]{}
\newcommand{\yj}[1]{}
\newcommand{\yibiao}[1]{}
\newcommand{\del}[1]{}
}{
\newcommand{\mytodogreen}[1]{\textcolor{mygreen}{\ding{46}~{\sf}~#1}}
\newcommand{\yibiao}[1]{\mytodogreen{[yibiao: #1]}}
\newcommand{\civi}[1]{\textcolor{red}{[Ming: #1]}}
\newcommand{\yj}[1]{\textcolor{blue}{[yj: #1]}}
\newcommand{\del}[1]{\textcolor{red}{\sout{#1}}}
}

\setcopyright{acmcopyright}
\copyrightyear{2018}
\acmYear{2018}
\acmDOI{XXXXXXX.XXXXXXX}

\acmJournal{JACM}
\acmVolume{37}
\acmNumber{4}
\acmArticle{111}
\acmMonth{8}

\begin{document}

\newcommand{\commitsFilesZ}{
    \pgfplotstableread[row sep=\\,col sep=&]{
        interval & count \\
        1&185\\
        2&81\\
        3&46\\
        4&19\\
        5&22\\
        6&7\\
        7&4\\
        8&5\\
        9&1\\
        11&3\\
        13&1\\
        17&1\\
        55&1\\
        65&1\\
        }\mydata    
\begin{tikzpicture}
    \tikzset{every node}=[font=\tiny\sffamily]
    \begin{axis}[
        ybar,
        bar width=.20cm,
        width=.82\textwidth,
        height=4.0cm,
        legend style={at={(0.5,1)},
        anchor=north,legend columns=-1},
        symbolic x coords={1,2,3,4,5,6,7,8,9,11,13,17,55,65},
        xtick pos=left,
        ytick pos=left,
        xtick=data,
        enlarge x limits={abs=0.4cm},
        nodes near coords,
        nodes near coords align={vertical},
        ymin=0,ymax=250,
        ylabel={\#Commits},
        xlabel={\#File changes},
        label style={font=\footnotesize},
        ]
        \addplot[fill=black!80] table[x=interval,y=count]{\mydata};
    \end{axis}
\end{tikzpicture}
}

\newcommand{\commitsFilesCVC}{
    \pgfplotstableread[row sep=\\,col sep=&]{
        interval & count \\
        1&59\\
        2&24\\
        3&6\\
        4&6\\
        5&3\\
        6&1\\
        13&1\\
        18&1\\
        }\mydata
    \begin{tikzpicture}
        \tikzset{every node}=[font=\tiny\sffamily]
        \begin{axis}[
            ybar,
            bar width=.20cm,
            width=1\textwidth,
            height=3cm,
            legend style={at={(0.5,1)},
            anchor=north,legend columns=-1},
            symbolic x coords={1,2,3,4,5,6,13,18},
            xtick pos=left,
            ytick pos=left,
            xtick=data,
            enlarge x limits={abs=0.4cm},
            nodes near coords,
            nodes near coords align={vertical},
            ymin=0,ymax=82,
            ylabel={\#Commits},
            xlabel={\#File changes},
            label style={font=\footnotesize},
            ]
            \addplot[fill=white] table[x=interval,y=count]{\mydata};
        \end{axis}
    \end{tikzpicture}
}

\newcommand{\releaseZ}{
    \pgfplotstableread[row sep=\\,col sep=&]{
        interval & count \\
        4.8.1&3\\
        4.9&6\\
        4.10&6\\
        4.11.0&6\\
        4.12.0&8\\
        4.13.0&11\\
        trunk&25\\
        }\mydata
    \begin{tikzpicture}
        \tikzset{every node}=[font=\scriptsize\sffamily]
        \begin{axis}[
            ybar,
            bar width=.3cm,
            width=0.9\textwidth,
            height=4cm,
            xticklabel style={rotate=40},
            legend style={at={(0.5,1)},
                anchor=north,legend columns=-1},
            symbolic x coords={4.8.1,4.9,4.10,4.11.0,4.12.0,4.13.0,trunk},
            xtick pos=left,
            ytick=\empty,
            xtick=data,
            enlarge x limits={abs=0.5cm},
            nodes near coords,
            nodes near coords align={vertical},
            ymin=0,ymax=35,
            label style={font=\footnotesize},
            ]
            \addplot[fill=black!80] table[x=interval,y=count]{\mydata};
        \end{axis}
    \end{tikzpicture}
}

\newcommand{\releaseCVC}{
    \pgfplotstableread[row sep=\\,col sep=&]{
        interval & count \\
        0.0.2&1\\
        0.0.11&2\\
        1.0.1&4\\
        1.1.0&5\\
        1.2.0&8\\
        trunk&18\\
        }\mydata
    \begin{tikzpicture}
        \tikzset{every node}=[font=\scriptsize\sffamily]
        \begin{axis}[
            ybar,
            bar width=.3cm,
            width=0.9\textwidth,
            height=4cm,
            xticklabel style={rotate=40},
            legend style={at={(0.5,1)},
                anchor=north,legend columns=-1},
            symbolic x coords={0.0.2,0.0.11,1.0.1,1.1.0,1.2.0,trunk},
            xtick pos=left,
            ytick=\empty,
            xtick=data,
            enlarge x limits={abs=0.5cm},
            nodes near coords,
            nodes near coords align={vertical},
            ymin=0,ymax=24,
            label style={font=\footnotesize},
            ]
            \addplot[fill=white] table[x=interval,y=count]{\mydata};
        \end{axis}
    \end{tikzpicture}
}

\newcommand{\comparison}{
    \pgfplotstableread[row sep=\\,col sep=&]{
        interval & count \\
        OpFuzz&3\\
        TypeFuzz&6\\
        STORM&3\\
        YinYang&1\\
        HistFuzz&11\\
        }\mydata
    \begin{tikzpicture}
        \tikzset{every node}=[font=\tiny\sffamily]
        \begin{axis}[
            ybar,
            bar width=.3cm,
            width=\textwidth,
            height=4.5cm,
            xticklabel style={rotate=0},
            legend style={at={(0.5,1)},
                anchor=north,legend columns=-1},
            symbolic x coords={OpFuzz,TypeFuzz,STORM,YinYang,HistFuzz},
            xtick pos=left,
            ytick=\empty,
            xtick=data,
            enlarge x limits={abs=0.5cm},
            nodes near coords,
            nodes near coords align={vertical},
            ymin=0,ymax=14,
            label style={font=\footnotesize},
            ]
            \addplot[fill=white] table[x=interval,y=count]{\mydata};
        \end{axis}
    \end{tikzpicture}
}

\newcommand{\component}{
\pgfplotsset{
tick label style={font=\tiny},
}
\begin{tikzpicture}[scale=1.0]
\begin{axis}[
    ymin=0.2,
    height=3.5cm,
    width=6cm,
    xtick pos=left,
    ytick pos=left,
    enlarge x limits={abs=0.3cm},
    every node near coord/.append style={font=\tiny\sffamily},
    xtick={0,12,24,36,48,60,72,84,96,108,120},
    ytick={2,4,6,8,10,12,14},
    label style={font=\footnotesize},
    ]
\addplot[line width=1pt,black!72,mark options={mark size=1pt},smooth] coordinates {
    (0,0)
    (0.5,2)
    (1,3)
    (1.5,4)
    (3,5)
    (12,6)
    (83.5,7)
    (120,7)
};
\addplot[line width=1pt,red!80,,mark options={mark size=0.8pt},smooth,dash pattern=on 3pt off 2pt on 1pt] coordinates {
    (0,0)
    (0.5,2)
    (1,3)
    (3,4)
    (6,5)
    (26,6)
    (30,7)
    (36.5,8)
    (69.5,9)
    (120,9)
};
\addplot[line width=1pt,blue!80,mark options={mark size=1.5pt},smooth, dashed] coordinates {
    (0,0)
    (0.5,3)
    (1,4)
    (1.5,5)
    (2,6)
    (6,7)
    (17,8)
    (40,9)
    (82,10)
    (120,10)
};
\end{axis}
\end{tikzpicture}
}

\newcommand{\reproduce}{
\pgfplotsset{
tick label style={font=\tiny},
}
\begin{tikzpicture}[scale=1.0]
\begin{axis}[
    ymin=0.2,
    height=4cm,
    width=7cm,
    xtick pos=left,
    ytick pos=left,
    enlarge x limits={abs=0.4cm},
    every node near coord/.append style={font=\tiny},
    xtick={1,2,3,4,5,6,7,8,9,10},
    ytick={12,24,36,48,60,72,84,96},
    ]
\addplot[line width=0.8pt,orange,mark options={mark size=0.9pt},mark=*] coordinates {
    (1,0.4)
    (2,0.54)
    (3,96)
    (4,4.1)
    (5,11.38)
    (6,21.3)
    (7,96)
    (8,4.68)
    (9,96)
    (10,96)
};
\addplot[line width=0.8pt,cyan!70!black,mark options={mark size=0.9pt},mark=*,dashed] coordinates {
    (1,0.2)
    (2,0.46)
    (3,48.95)
    (4,6.8)
    (5,8)
    (6,9.1)
    (7,16.8)
    (8,0.7)
    (9,82.1)
    (10,95)
};

\end{axis}
\end{tikzpicture}
}

\newcommand{\typetable}{
\begin{figure}[t]
    \renewcommand{\arraystretch}{1.2}
    \setlength{\tabcolsep}{3.5pt}
    \centering
    \footnotesize
    \begin{tabular}{ll}   
        \toprule    
        \textbf{Conversion Function Symbol}  &\textbf{Description} \\    
        \midrule          
        \texttt{ite}  & Convert \texttt{Bool} to any sort \\  
        \texttt{is_int} & Convert \texttt{Int} or \texttt{Real} to \texttt{Bool} \\ 
        \texttt{to_real} & Convert \texttt{Int} to \texttt{Real} \\       
        \texttt{str.from_int,str.from_code} & Convert \texttt{Int} to \texttt{Str} \\   
        \texttt{int2bv} & Convert \texttt{Int} to \texttt{BitVec}  \\   
        \texttt{to_int} & Convert \texttt{Real} to \texttt{Int} \\
        \texttt{str.is_digit} & Convert \texttt{Str} to \texttt{Bool} \\    
        \texttt{str.to_int,str.to_code} & Convert \texttt{Str} to \texttt{Int} \\
        \texttt{bv2nat} & Convert \texttt{BitVec} to \texttt{Int} \\

        \bottomrule     
    \end{tabular}
\label{typetable}
\caption{Conversion function symbols defined in SMT-LIB. 
    \label{fig:op-type-table}} 
\end{figure}
}

\newcommand{\bugcounttable}{
{\small
\setlength{\tabcolsep}{0.8em}
\renewcommand{\arraystretch}{1.21}
\begin{tabular}{lrr|r}
\toprule
\textbf{Status} &  \textbf{Z3} &  \textbf{cvc5} &  \textbf{Total} \\
\midrule
Reported  &  27 &  18 &     45 \\
Confirmed &  25 &   18 &     43 \\
Fixed     &  24 &   16 &    40 \\
Duplicate &  2 &    0 &     2  \\
\bottomrule
\end{tabular}
}
}

\newcommand{\bugtypetable}{
{\small
\setlength{\tabcolsep}{0.8em}
\renewcommand{\arraystretch}{1.49}
\begin{tabular}{lrr|r}
\toprule
\textbf{Type} &  \textbf{Z3} &  \textbf{cvc5}  &  \textbf{Total} \\
\midrule
Crash         &  20 &     15  &  35  \\ 
Invalid model  &  4 &      2 &   6 \\ 
Soundness &   3 &      1  &   4 \\ 
\bottomrule
\end{tabular}
}
}

\newcommand{\validproportion}{
\definecolor{Color2}{RGB}{200,36,35}
\definecolor{Color1}{RGB}{40,120,181}
\begin{tikzpicture}
\pgfplotsset{
tick label style={font=\footnotesize},
}
\begin{axis}[
height=5cm,
width=8cm,
ymin=0.1,
ymax=1,
xmin=0.05,
xmax=1.05,
xtick={0.1,0.2,0.3,0.4,0.5,0.6,0.7,0.8,0.9,1.0},
ytick={0.1,0.2,0.3,0.4,0.5,0.6,0.7,0.8,0.9,1.0},
xlabel={\small \textbf{Temperature}},
ylabel={\small \textbf{Proportion}},
legend style={at={(0.97,0.97)}, anchor=north east},
grid=major,
grid style={dashed,gray!30},
every axis plot/.append style={thick},
]
\addplot[
color=Color2,
mark=*,
mark size=1.5pt,
]
coordinates {
(0.1,0.76)
(0.2,0.86)
(0.3,0.77)
(0.4,0.79)
(0.5,0.82)
(0.6,0.79)
(0.7,0.80)
(0.8,0.76)
(0.9,0.75)
(1.0,0.74)
};

\addplot[
color=Color1, %
mark=square*, %
mark size=1.5pt,
]
coordinates {
(0.1,0.57) %
(0.2,0.67)
(0.3,0.61)
(0.4,0.59)
(0.5,0.63)
(0.6,0.69)
(0.7,0.63)
(0.8,0.62)
(0.9,0.62)
(1.0,0.60)
};

\end{axis}
\end{tikzpicture}
}

\newcommand{\diffplot}{
\begin{tikzpicture}
\pgfplotsset{
    tick label style={font=\footnotesize},
}
\begin{axis}[
    height=4cm,
    width=7cm,
    ymin=0,
    ymax=4,
    xmin=0,
    xmax=11,
    xtick={1,2,3,4,5,6,7,8,9,10},
    ytick={0,0.5,1,1.5,2,2.5,3,3.5},
    xlabel={\small \textbf{Iteration}},
    ylabel={\small \textbf{Average Diff}},
    legend style={at={(0.97,0.03)}, anchor=south east},
    grid=major,
    grid style={dashed,gray!30},
    every axis plot/.append style={thick},
]
\addplot[
    color=black,
    mark=*,
    mark size=1.5pt,
]
coordinates {
    (0,0)
    (1,0.4938543598367193)
    (2,0.776657504186989)
    (3,1.0911566534235457)
    (4,1.3688284920380223)
    (5,1.6170494222840548)
    (6,2.0495838598051965)
    (7,2.3644366803914223)
    (8,2.706227309649664)
    (9,2.998747987240373)
    (10,3.2501485958999914)
};

\end{axis}
\end{tikzpicture}
}

\newcommand{\proportion}{
\pgfplotstableread[row sep=\\,col sep=&]{
interval & count \\
YinYang & 46 \\
OpFuzz & 100 \\
TypeFuzz & 88.3 \\
HistFuzz & 51 \\
LaST & 59 \\
SeaWall-zs & 54 \\
SeaWall-fs & 67 \\
}\mydata
\begin{tikzpicture}
\tikzset{every node/.append style={font=\scriptsize\sffamily}}
\begin{axis}[
ybar,
bar width=0.5cm, %
width=0.5\textwidth,
height=4.8cm,
xticklabel style={rotate=40},
legend style={at={(0.5,1)},anchor=north,legend columns=-1},
symbolic x coords={YinYang,OpFuzz,TypeFuzz,HistFuzz,LaST,SeaWall-zs,SeaWall-fs, Similarly},
xtick pos=left,
ytick pos=left,
ylabel={Proportion},
ytick distance=20, %
xtick=data,
enlarge x limits={abs=1cm},
ymin=0,ymax=110,
label style={font=\footnotesize},
]

\addplot[fill=gray!70] table[x=interval,y=count]{\mydata};
\addplot[fill=brown!70] coordinates {(Similarly, 46)};

\end{axis}
\end{tikzpicture}
}

\title{Once4All: Skeleton-Guided SMT Solver Fuzzing with LLM-Synthesized Generators}

\author{Maolin Sun}
\orcid{0000-0001-5617-2205}
\affiliation{%
  \institution{State Key Laboratory for Novel Software Technology, Nanjing University}
  \city{Nanjing}
  \country{China}}
  \email{merlin@smail.nju.edu.cn}

\author{Yibiao Yang}
\orcid{0000-0003-1153-2013}
\affiliation{%
  \institution{State Key Laboratory for Novel Software Technology,  Nanjing University}
  \city{Nanjing}
  \country{China}}
\email{yangyibiao@nju.edu.cn}
\authornote{Corresponding author.}

\author{Yuming Zhou}
\orcid{0000-0002-4645-2526}
\affiliation{%
  \institution{State Key Laboratory for Novel Software Technology, Nanjing University}
  \city{Nanjing}
  \country{China}}
\email{zhouyuming@nju.edu.cn}

\renewcommand{\shortauthors}{Maolin Sun, Yibiao Yang, and Yuming Zhou}

\begin{abstract}
Satisfiability Modulo Theory (SMT) solvers are foundational to modern systems and programming languages research, providing the foundation for tasks like symbolic execution and automated verification.
Because these solvers sit on the critical path, their correctness is essential, and high-quality test formulas are key to uncovering bugs. 
However, while prior testing techniques performed well on earlier solver versions, they struggle to keep pace with rapidly evolving features.
Recent approaches based on Large Language Models (LLMs) show promise in exploring advanced solver capabilities, but two obstacles remain: nearly half of the generated formulas are syntactically invalid, and iterative interactions with LLMs introduce substantial computational overhead.
In this study, we present \tool, a novel LLM-assisted fuzzing framework that addresses both issues by shifting from direct formula generation to the synthesis of generators for reusable terms (\ie~logical expressions) .
Specifically, \tool~uses LLMs to (1) automatically extract context-free grammars (CFGs) for SMT theories, including solver-specific extensions, from documentation, and (2) synthesize composable Boolean term generators that adhere to these grammars.
During fuzzing, \tool~populates structural skeletons derived from existing formulas with the terms iteratively produced by the LLM-synthesized generators. 
This design ensures syntactic validity while promoting semantic diversity.
Notably, \tool~requires only \textbf{one-time LLM interaction investment}, dramatically reducing runtime cost. 
We evaluated \tool~on two leading SMT solvers: Z3 and cvc5. Our experiments show that \tool~has identified \confirmed~confirmed bugs, \fixed~of which have already been fixed by developers.

\end{abstract}



\maketitle

\section{Introduction}
\label{sec:intro}

Satisfiability Modulo Theory (SMT) solvers are essential systems for automatically determining the satisfiability of logical formulas with respect to background theories, such as arithmetic, bit-vectors, and strings~\cite{barrett2018satisfiability}. 
They play a critical role across multiple domains, including computer systems~\cite{DBLP:conf/sosp/JiaPTWZA19,DBLP:conf/sosp/NelsonSZJBTW17,DBLP:conf/osdi/HawblitzelHLNPZZ14,DBLP:conf/osdi/HanceLHHJP20}, computer networks~\cite{DBLP:conf/asplos/HanWQSXWZYL24,DBLP:conf/nsdi/LopesBGJV15,DBLP:conf/nsdi/El-HassanyTVV18,DBLP:conf/nsdi/KakarlaBMV22,DBLP:conf/sigcomm/TangKBZBMTV21}, and programming languages~\cite{DBLP:conf/popl/LiAKGC14,DBLP:conf/cav/KangLT19,DBLP:conf/asplos/LiuLRTLPZ23}. 
For instance, the renowned symbolic execution engine KLEE~\cite{cadar2008klee} relies on SMT solvers to check the feasibility of program paths.
However, bugs hidden in SMT solvers can  mislead client applications, resulting in serious consequences and even security vulnerabilities~\cite{CVE19725}.
Thus, ensuring the reliability of SMT solvers is of critical importance.

To this end, various approaches have been proposed for testing SMT solvers, which can be broadly classified into three categories: (1)~\textit{generation-based}~\cite{brummayer2009fuzzing,DBLP:conf/cav/NiemetzPB22,DBLP:journals/pacmpl/Winterer024}, (2)~\textit{mutation-based }~\cite{winterer2020validating,mansur2020detecting,DBLP:journals/pacmpl/WintererZS20,DBLP:journals/pacmpl/ParkWZS21,DBLP:conf/icse/KimSO23,sunvalidating}, and (3)~\textit{LLM-based}~\cite{sunsmt,DBLP:journals/corr/abs-2308-04748}.
Generation-based approaches typically generate formulas from scratch based on predefined generation rules or grammar.
For instance, ET~\cite{DBLP:journals/pacmpl/Winterer024} is a representative technique that enumerates formulas based on grammar.
Mutation-based approaches, on the other hand, involve mutating existing formulas to generate new ones.
One of the most advanced approaches in this category is HistFuzz~\cite{sunvalidating}, which utilizes historical bug-triggering formulas to generate new formulas.
Furthermore, researchers have explored LLM-based approaches inspired by the advancement of large language models (LLMs).
These approaches aim to generate SMT formulas directly using LLMs~\cite{sunsmt,DBLP:journals/corr/abs-2308-04748}.
For instance, LaST~\cite{sunsmt} retrained an LLM, aiming to generate effective test formulas from scratch.
Fuzz4All~\cite{DBLP:journals/corr/abs-2308-04748} is a universal framework that is applicable to various software systems, including SMT solvers.
It can generate test inputs for SMT solvers by prompting the generation LLM with the prompt generated by another LLM.

Despite notable progress, existing SMT solver fuzzing approaches still face several challenges. 
Generation-based techniques often struggle to produce diverse formulas because they rely on rigid generative rules, and their from-scratch generation process makes it costly to reach deeper solver states. 
In comparison, mutation-based approaches can leverage seed formulas to more efficiently explore deeper execution paths. 
However, their effectiveness still depends heavily on the design of mutation strategies~\cite{DBLP:journals/pacmpl/ParkWZS21,sunsmt}. 
For instance, OpFuzz~\cite{DBLP:journals/pacmpl/WintererZS20} only mutates the operators in seed formulas, which may restrict the diversity of generated formulas.
More recently, LLM-based techniques directly prompt large language models to generate complete SMT formulas. 
While attractive in principle, such approaches often produce inputs that are syntactically or semantically invalid (approximately  50\%)~\cite{sunsmt,DBLP:journals/corr/abs-2308-04748}, sometimes lead to nonsensical bug reports\footnote{\url{https://github.com/Z3Prover/z3/issues/6818}}, and incur substantial computational cost due to heavy interaction with LLMs. 
Meanwhile, the SMT-LIB standard~\cite{SMTLIB}, the primary input language for SMT solvers, continues to evolve.
The recent release of version 2.7 and the planned introduction of SMT-LIBv3 reflect the community's ongoing efforts to support richer language features. 
Concurrently, SMT solvers like cvc5 have introduced extended, solver-specific theories~\cite{cvc5extend} to better express real-world constraints. 
Collectively, these developments increase the demand for fuzzing techniques that can adapt to evolving changes for better testing emerging solver capabilities.

\mypara{Our Approach.}
In this study, we propose \tool, a novel SMT solver fuzzing framework that unifies the strengths of mutation-based fuzzing with the advanced capabilities of LLMs.
While mutation-based fuzzing techniques
achieve high testing throughput and deep state exploration by mutating existing test cases, adapting them to support newly introduced or evolving SMT features typically requires substantial manual effort.
\tool~overcomes this limitation by leveraging LLMs' contextual understanding and program synthesis abilities to enable rapid, automated adaptation to changes in solver input language. 
Specifically, \tool~includes two key phases: (1)~\textbf{\textit{LLM-assisted generator construction}} and (2)~\textbf{\textit{skeleton-guided mutation}}.
In the first phase, \tool~prompts LLMs to extract and summarize the context-free grammars (CFGs) of SMT-LIB theories, drawing from the official documentation~\cite{SMTLIB}. We also include grammar for solver-specific extensions, such as the Bag and Set theories in cvc5~\cite{cvc5extend}.
Based on these CFGs, LLMs then synthesize reusable generators that emit 
Boolean terms (\ie~logical expressions) conforming to the theories.
This process is a \textit{one-time investment}, only requiring repetition when grammar evolves. 
Note that this phase is essential because no existing resource provides off-the-shelf grammars or generators that fully cover both the latest SMT-LIB standard and diverse solver-specific extensions~\cite{DBLP:journals/pacmpl/Winterer024}.
Particularly, many extensions are only informally documented~\cite{cvc5extend}, making manual construction difficult, whereas LLMs can readily interpret and adapt to such documentation.
The second phase employs these generators within a skeleton-guided mutation procedure. 
The skeletons are derived by removing small Boolean terms from seed formulas~\cite{sunvalidating}.  
The terms generated by the LLM-implemented generators are then synthesized into the skeletons, producing novel test formulas.
Skeletons help target deeper solver logic while also complementing gaps in SMT-LIB documentation.
For example, important symbols like quantifiers are not specified in theory documentation, limiting what a purely grammar-based generator can produce. 
By utilizing existing skeletons, \tool\ expands the expressiveness of generated formulas. 
\tool~finally performs differential testing across
multiple solvers
to detect behavioral divergences indicative of bugs.

We have implemented \tool~as a practical and extensible fuzzing tool for SMT solvers.
To evaluate the effectiveness of \tool, we applied it to the leading SMT solvers, \ie~Z3~\cite{de2008z3} and cvc5.
Ultimately, \tool~has successfully identified \reported~real bugs in Z3 and cvc5, of which \confirmed~were confirmed, and \fixed~of those confirmed bugs were fixed by the developers.
Notably, \tool~exhibits effectiveness in identifying bugs across various theories, including the recently introduced theories and solver-specific extensions.
Besides, our experiments also exhibit that \tool~outperforms advanced SMT solver fuzzers in terms of achieved code coverage and effectiveness.

\mypara{Contributions.} We make the following major contributions:

 \begin{itemize}[leftmargin=*,label={$\star$}]
    \item \mypara{Novelty:}~We propose a novel approach for fuzzing SMT solvers, combining the strengths of mutation-based fuzzing with the capabilities of LLMs. By strategically using LLMs, we enhance the generation of test formulas, allowing our approach to adapt more effectively to the evolving features of SMT solvers.
    \item \mypara{Significance:}~Our work focuses on SMT solvers, which are fundamental components in formal verification and program analysis. Improving their robustness is of paramount importance. Furthermore, we present a fresh perspective on applying LLMs in software testing, a paradigm that can be extended to other software systems beyond SMT solvers. 
    \item \mypara{Practicality:}~We implement our idea as a practical tool \tool. \tool~has led to \reported~bug reports for the leading SMT solvers Z3 and cvc5, of which \confirmed~are confirmed and \fixed~are fixed by developers. Experimental results also demonstrate that \tool~outperforms the state-of-the-art SMT solver fuzzers in terms of achieved code coverage. 
 \end{itemize}

\mypara{Paper Organization.} 
The remainder of this paper is organized as follows. 
Section~\ref{sec:motivation} provides an overview of the necessary background and motivates our approach.
Section~\ref{sec:approach} formalizes our approach and describes the implementation of~\tool. 
Next, we elaborate on our extensive evaluation in detail in Section~\ref{sec:evaluation}, and Section~\ref{sec:discussion} is dedicated to in-depth discussions on the results. 
Finally, we survey related work in Section~\ref{sec:related-work}, and the conclusion is in Section~\ref{sec:conclusion}.

\section{Background and Motivation}
\label{sec:motivation}

In this section, we first overview the SMT-LIB language and the standard for describing SMT formulas. 
Then, we motivate our technique by illustrating a real-world bug in the cvc5 SMT solver.

\subsection{SMT-LIB Language}

SMT-LIB serves as the standard input language for SMT solvers, providing a unified framework for expressing logical formulas and background theories. Developed by the SMT-LIB initiative, it defines a formal syntax and semantics for declaring symbols, asserting constraints, and querying satisfiability. 
The core commands in SMT-LIB include \texttt{declare-fun} for symbol declarations, \texttt{assert} for adding logical constraints, and \texttt{check-sat} for determining satisfiability. 
Note that variables can be treated as functions with no arguments.
For example, the statement \texttt{(declare-fun x () Int)} introduces an integer variable \texttt{x}, and \texttt{(assert (and (< x 3) (> y 1)))} requires that \texttt{x} is less than 3 and \texttt{y} is greater than 1.
The solver returns \texttt{sat} if all assertions can be satisfied simultaneously, or \texttt{unsat} otherwise.
SMT-LIB is continuously evolving to support more complex and real-world scenarios~\cite{SMTnews}. 
Recent versions, such as version 2.7~\cite{BarFT-RR-25} (released in 2025), have introduced advanced features like polymorphic sorts, higher-order functions, and wildcards in pattern matching. These extensions significantly enhance the expressiveness of SMT solver inputs, allowing for the modeling of more intricate problems without compromising backward compatibility or the core principles of the language.

\subsection{Motivation}
The importance of SMT solvers in the software engineering community has led to the development of numerous fuzzing techniques aimed at ensuring their correctness and robustness.
These methods have successfully uncovered many bugs~\cite{DBLP:conf/cav/NiemetzPB22,DBLP:journals/pacmpl/WintererZS20,DBLP:journals/pacmpl/ParkWZS21,sunvalidating}.
However, most existing approaches rely on manually designed input-generation strategies. 
As the SMT-LIB standard evolves and solvers introduce new features, such static strategies often struggle to keep pace. 
This limitation reflects a broader challenge in software testing: as systems evolve, subtle bugs can emerge that existing methods may overlook, making adaptive testing approaches increasingly necessary~\cite{DBLP:journals/corr/abs-2308-04748}.
To illustrate, consider a real bug in the cvc5 solver, shown in Figure~\ref{fig:motivate}. 
This bug occurred in cvc5's handling of the sequence theory, which was extended to support comprehensive operations for modeling real-world sequences. 
In this example, \texttt{seq.rev} reverses a sequence, \texttt{seq.len} computes its length, and \texttt{seq.nth} retrieves an element at a specific position (\ie~$n$-th).
The formula in Figure~\ref{fig:motivate} triggered unexpected behavior in cvc5. 
Specifically, the sequence-related term could not be evaluated to a concrete constant, causing the solver to incorrectly reject a valid model. 
This example illustrates a broader issue: specialized and newly added features often involve intricate logic that generic fuzzing techniques, originally designed for established theories, fail to test adequately.
Fortunately, recent advances in LLMs offer a promising way forward. 
In particular, LLMs demonstrate strong capabilities in understanding both code and documentation~\cite{DBLP:journals/corr/abs-2308-04748}, suggesting they could help automate the tracking and testing of newly integrated solver features without heavy human effort. 

\begin{observationbox}
\textit{Beyond established theories, newly introduced extensions and solver-specific features are also sources of bugs and therefore demand targeted testing.}
    
\end{observationbox}

Furthermore, the logical structure of the input formula, \ie~its skeleton, can be critical in exposing such bugs. 
In the example shown in Figure~\ref{fig:motivate}, developers noted that the quantifier was also a critical factor in triggering the bug, even though it was semantically largely irrelevant to the problematic core expression.\footnote{\url{https://github.com/cvc5/cvc5/pull/11930\#discussion_r2104698083}}
In other words, the bug cannot be reproduced without the presence of the quantifier \texttt{exists ((f Int))}, despite its lack of direct interaction with the assertion's body.
Its presence altered the solver's internal execution path, revealing the bug in model evaluation.
This demonstrates that seemingly minor structural elements within a formula can significantly influence a solver's behavior and are critical for uncovering bugs, a phenomenon also observed in other studies~\cite{sunvalidating}.
Therefore, this leads to the following observation.
\begin{observationbox}

\textit{The skeleton of the input formula, including quantifiers and logical connectives, can profoundly influence solver behavior and is crucial for revealing bugs.}
\end{observationbox}

In practice, this bug was discovered by our proposed technique, \tool. 
Specifically, \tool~extracts skeletons from existing formulas by removing their atomic sub-formulas.
In this example, the skeleton is \texttt{(exists ((f Int))} \colorbox{gray!40!white}{\texttt{<placeholder>}}\texttt{)}.
Subsequently, \tool~fills in the placeholders with newly generated formulas across different theories.
We will elaborate on the details in Section~\ref{sec:approach}.

\begin{figure}[t]
    \centering
\begin{lstlisting}[basicstyle=\footnotesize\ttfamily]
(declare-fun s () (Seq Int))
(assert (exists ((f Int)) 
 (distinct (seq.len (seq.rev s)) 
 (seq.nth (as seq.empty (Seq Int)) (div 0 0)))))
(check-sat)
\end{lstlisting}
    \caption{A bug-revealing formula involving the \texttt{Seq} theory extended in cvc5. \href{https://github.com/cvc5/cvc5/issues/11924}{\#11924}}
    \label{fig:motivate}
\end{figure}

\section{Approach}
\label{sec:approach}
\begin{figure*}[ht]
    \centering
\includegraphics[width=0.9\linewidth]{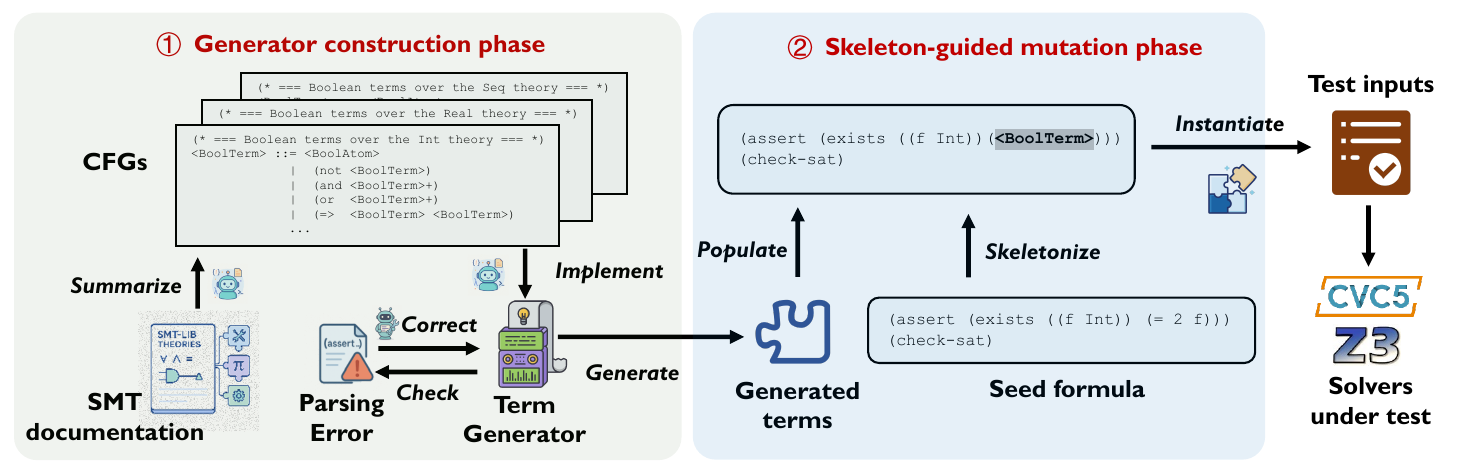}
	\caption{Overview of \tool. \label{fig:overview}}
\end{figure*}

In this section, we provide an overview of our proposed approach, \tool, followed by a detailed explanation. 

\subsection{Overview}

This study introduces a novel approach called \tool, illustrated in Figure~\ref{fig:overview}. 
\tool's workflow is structured into two primary phases: (1) LLM-assisted generator construction and (2) skeleton-guided mutation.
The first phase focuses on building robust test case generators. 
We begin by gathering comprehensive documentation on various SMT theories from official sources, including the SMT-LIB standard and leading SMT solvers like Z3 and cvc5. 
Subsequently, we leverage LLMs to automatically summarize the context-free grammar (CFG) for each theory. 
With these grammars, we then instruct LLMs to implement generators capable of producing Boolean terms for each theory. These generators are designed to create Boolean terms that adhere strictly to their respective CFGs, include necessary declarations such as \texttt{declare-fun} and \texttt{declare-datatypes}, and are fully compatible with the SMT-LIB specification. This initial phase is a one-time setup process, requiring updates only if new theories or features are introduced to SMT solvers.
Upon completion of the generator construction, we advance to the second phase: skeleton-guided mutation. 
In this phase, \tool~takes existing formulas, such as those found in official SMT-LIB benchmarks, and extracts their skeletons by removing specific Boolean terms. 
\tool~then calls upon the generators developed in the first phase to produce new Boolean terms. 
These newly generated terms are then integrated into the extracted skeletons, forming novel test formulas. 
Finally, \tool~employs differential testing by feeding the generated test formulas to multiple SMT solvers and analyzing their behaviors to identify discrepancies.

\newcommand\mycommfont[1]{\footnotesize\ttfamily\textcolor{blue}{#1}}
\SetCommentSty{mycommfont}
\normalem
\setlength{\textfloatsep}{0.5cm}

\begin{algorithm}[t]
\small
\SetKwInOut{Input}{Input}
\SetKwInOut{Output}{Output}

\Input{Set of documentation for SMT theories $Doc_T$, Large Language Model $LM$}
\Output{Set of self-corrected generators $G$}
\SetKwFunction{FnSummarizeCFG}{SummarizeCFG}
\SetKwFunction{FnConstructPrompt}{ConstructPrompt}
\SetKwFunction{FnInvokeLLM}{InvokeLLM}
\SetKwFunction{FnCorrect}{Correct}
\SetKwFunction{FnRefine}{Refine}
\SetKwFunction{FnGenerate}{Generate}
\SetKwFunction{FnCheckErr}{CheckErr}
\SetKwFunction{FnParse}{Parse}
\SetKwProg{Fn}{Function}{}{end}
\SetKwProg{Fp}{Procedure}{}{end}
\DontPrintSemicolon

\BlankLine
\Fp{\texttt{GenConstruct($Doc_T$, $LM$)}}{
\BlankLine
$G \leftarrow \emptyset$\;

\ForEach{$doc \in Doc_T$}{
\tcc{Extract CFG from documentation}
$Prompt_1 \leftarrow$ \FnConstructPrompt($doc$) \;
$CFG \leftarrow $\FnSummarizeCFG($Prompt_1$, $LM$)\label{alg:cfg_extraction} \; 
\tcc{Use CFG to construct generator}
$Prompt_2 \leftarrow$ \FnConstructPrompt($CFG$) \;
$g_t \leftarrow$ \FnInvokeLLM($Prompt_2$) \label{alg:generator_creation}\; 
\tcc{Check and correct the generator}
$valid\_g_t \leftarrow$ \FnCorrect($g_t$, $LM$, $Solvers$) \label{alg:correction_call} \;
\BlankLine
$G \leftarrow G \cup $ $valid\_g_t$\;
}
}

\BlankLine

\Fn{\texttt{Correct($g$, $LM$, $Solvers$)} \label{alg:correction_procedure_start}}{

$iter \leftarrow 0$; $max\_iter \leftarrow 10$\;
$best\_g \leftarrow g$; $max\_valid \leftarrow 0$\;
$sample\_num \leftarrow 20$ \label{alg:sample_num}\;
\BlankLine
\While{$max\_valid < sample\_num$ \KwSty{and} $iter < max\_iter$}{ \label{alg:correction_while_loop}
$Errors \leftarrow \emptyset$\;
$iter \leftarrow iter + 1$\;
$terms \leftarrow$ \FnGenerate($g$, $sample\_num$) \label{alg:sample_generation}\; 
$valid\_cnt \leftarrow 0$\;

\ForEach{term $t \in terms$}{ \label{alg:parsing_and_error_check_loop_start}
$ParseRes \leftarrow$ \FnParse($t$)\label{alg:parsing_and_error_check}\; 
\If{\FnCheckErr($ParseRes$)}{ \label{alg:error_check_conditional}
$Errors \leftarrow Errors \cup ParseRes$\;
}\Else{
$valid\_cnt \leftarrow valid\_cnt + 1$\;
}
\label{alg:parsing_and_error_check_loop_end}} 

\If{$valid\_cnt > max\_valid$}{
$max\_valid \leftarrow valid\_cnt$\;
$best\_g \leftarrow g$\;
}

\If{$valid\_cnt < sample\_num$}{
$P \leftarrow$ \FnConstructPrompt($Errors$)\;
$g \leftarrow$ \FnRefine($P$, $LM$)\label{alg:generator_refinement}\; 
}
}
$g \leftarrow best\_g$ \label{alg:best_generator_update}\; 
\label{alg:correction_procedure_end}
}
    
\caption{\small LLM-Assisted Generator Construction}
\label{alg:gen_construct}
\end{algorithm}

\subsection{LLM-Assisted Generator Construction}

\begin{figure}[h!]
    \centering
\begin{subfigure}[b]{0.95\linewidth}
	\centering
	\includegraphics[width=\linewidth]{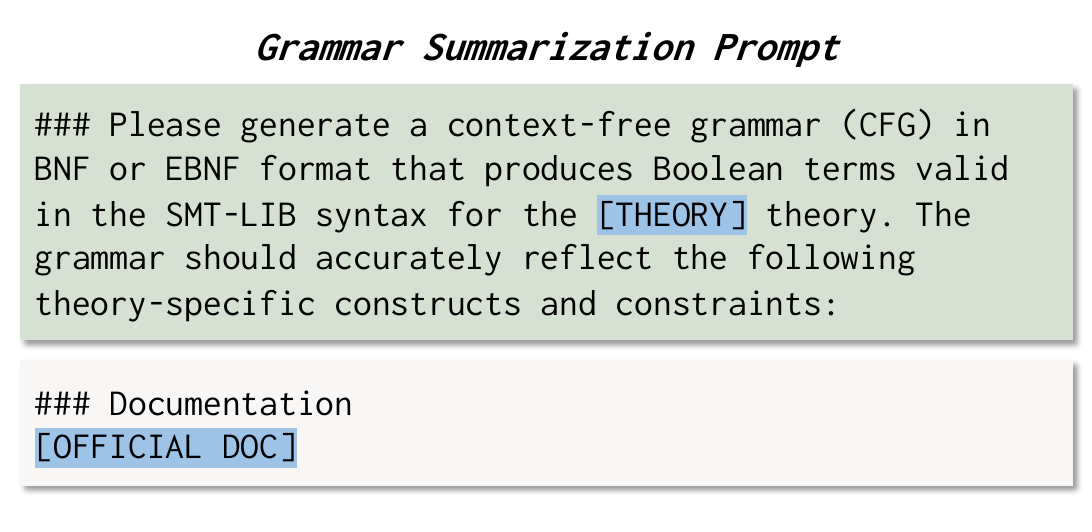}
	\caption{Prompt template for grammar summarization. \label{fig:prompt1}}
\end{subfigure}

\begin{subfigure}[b]{0.95\linewidth}
	\centering
	\includegraphics[width=\linewidth]{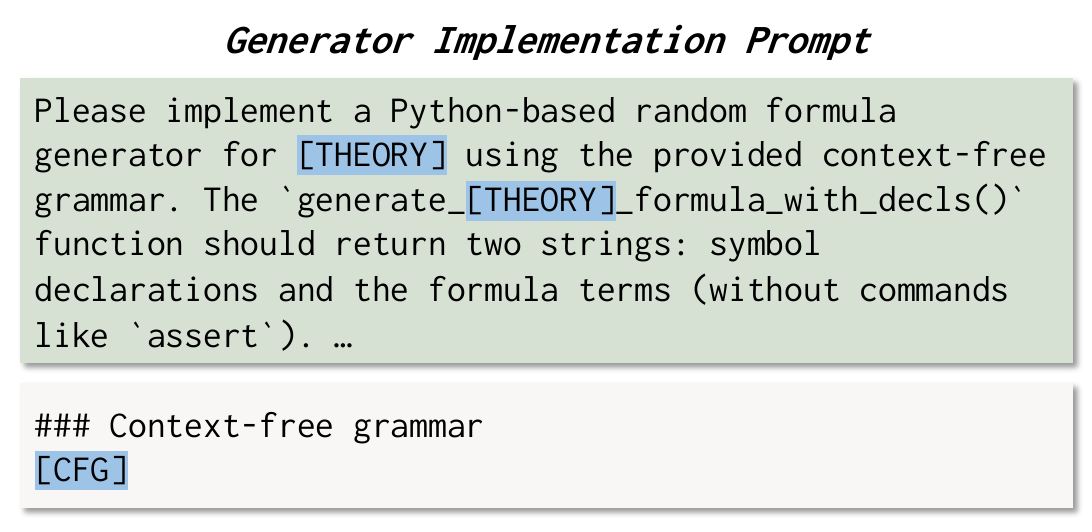}
	\caption{Prompt template for generator implementation. \label{fig:prompt2}}
\end{subfigure}

\begin{subfigure}[b]{0.95\linewidth}
	\centering
	\includegraphics[width=\linewidth]{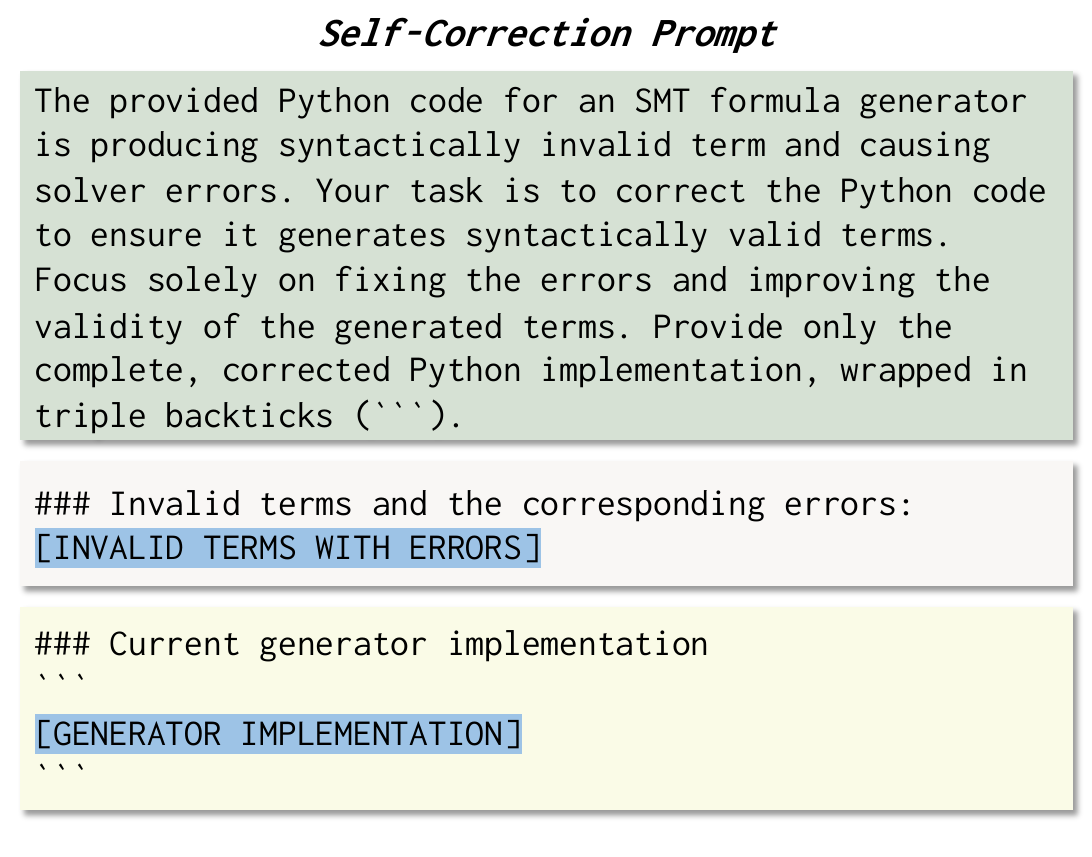}
	\caption{Prompt template for self-correction of generators. \label{fig:prompt3}}
\end{subfigure}

\caption{Prompt template used in LLM-assisted generator construction phase. \label{fig:prompt}}
\end{figure}

\mypara{Construction of Generators.}
The initial phase focuses on building generators for various SMT theories, as illustrated in Algorithm~\ref{alg:gen_construct}.
This automation is key to ensuring our fuzzing framework can readily adapt to the continuous evolution of SMT solvers.
Here, we divide this complex task into two sub-tasks: (1) summarizing context-free grammars and (2) implementing the generators, as shown in Algorithm~\ref{alg:gen_construct}.
Specifically, our approach commences by gathering comprehensive documentation for a range of SMT theories. 
This includes standard theories defined in SMT-LIB~\cite{SMTLIB} (\eg~\texttt{Ints}, \texttt{Reals}, \texttt{Bit-vectors}) and solver-specific extensions, such as Z3's \texttt{Unicode} theory~\cite{z3guide} and cvc5's \texttt{Set} theory~\cite{cvc5extend}.
Notably, many of these extensions are described only informally, making manual grammar construction difficult and prone to inconsistency.
Using this collected documentation, \tool~employs LLMs to automatically summarize the context-free grammar (CFG) for each theory.
This summarization (Line~\ref{alg:cfg_extraction}) is guided by a pre-defined prompt template, which is shown in Figure~\ref{fig:prompt1}.
Subsequently, these extracted CFGs form the basis for implementing the generators (Line~\ref{alg:generator_creation}). 
\tool~again leverages LLMs, providing them with the CFGs and a detailed instruction as shown in Figure~\ref{fig:prompt2}.
This step aims to produce generators capable of creating syntactically valid Boolean terms (\ie~logical expressions) for each theory. 
Specifically, the generated Boolean terms are expected to conform to the respective CFGs, include necessary SMT-LIB declarations (e.g., \texttt{declare-fun}, \texttt{declare-datatypes}), and adhere to the SMT-LIB specification.
This automated pipeline facilitates the creation of generators with a consistent interface, simplifying the future integration of new theories.

\mypara{Self-correction of Generators.}
Before these generators are employed in the fuzzing process, we incorporate a self-correction mechanism to ensure the syntactic and semantic validity of the Boolean terms they produce (Line~\ref{alg:correction_call}).
This mechanism is essential because the CFGs extracted from the documentation may not fully capture the intricacies of the SMT-LIB specification, potentially resulting in invalid terms.
For example, the CFGs for the BitVector theory may fail to enforce consistent bit-widths for operations that require operands of equal width, such as \texttt{bvadd} and \texttt{bvmul}.
The self-correction process begins by tasking a newly constructed generator with producing a small set of sample terms (Line~\ref{alg:sample_generation}).
The number of sample terms is set to 20, as shown in line~\ref{alg:sample_num}.
Each term is then augmented with necessary SMT-LIB commands and parsed by multiple SMT solvers, such as Z3 and cvc5.
A term is deemed valid if at least one solver parses it without error (Line~\ref{alg:parsing_and_error_check}).
If any of the sample terms are found to be invalid, \tool~initiates a refinement loop (Line~\ref{alg:correction_while_loop}). 
The parsing errors reported by the solvers are collected and used to construct a new prompt for the LLM based on the devised template (Figure~\ref{fig:prompt3}), instructing it to refine the generator's implementation (line~\ref{alg:generator_refinement}). 
Notably, if multiple terms are identified as invalid, we first use LLMs to distill and deduplicate the error messages, and then construct a prompt based on the distilled messages to guide the improvement of the generators.
This feedback-driven cycle is repeated until all sample terms are valid or a predefined maximum number of iterations (\eg 10) is reached. 
If the latter occurs, the version of the generator that produced the highest number of valid expressions across the iterations is retained (line~\ref{alg:best_generator_update}). 
This self-correction mechanism is designed to improve the quality of the generated terms, thereby enhancing the overall effectiveness of \tool. 
With these robust generators, \tool~is well-equipped for the subsequent phase of skeleton-guided fuzzing.
Notably, the self-correction mechanism cannot guarantee that the generators produce only well-formed terms; however, a small proportion of ill-formed terms is tolerable in the context of fuzzing.

\begin{algorithm}[t]
	\small 
    \SetKwInOut{Input}{Input}
    \SetKwInOut{Output}{Output}
    \SetKwFunction{FnRandomSelect}{RandomSelect}
    \SetKwFunction{FnSkeletonize}{Skeletonize}
    \SetKwFunction{FnSynthesize}{Synthesize}
    \SetKwFunction{FnGenerate}{Generate}
    \SetKwFunction{FnValidate}{Validate}
    \SetKwFunction{FnDiscrepancy}{Discrepancy}
    \SetKwProg{Fp}{Procedure}{}{end}
    \DontPrintSemicolon

    \Input{Generators $G$, Seed formulas $Seeds$}
    \Output{Bugs List $Bugs$}

    \BlankLine
    \Fp{\texttt{Fuzz($G$, $Seeds$, $Solvers$)}}{
        $Bugs \leftarrow \emptyset$ \label{alg:fuzz:init_bugs}
        \BlankLine
        \While{\textit{no termination by user}}{ \label{alg:fuzz:main_loop}
            $f \leftarrow \FnRandomSelect{Seeds}$ \label{alg:fuzz:select_seed}
            \BlankLine
            \Repeat{mutation termination criterion is met}{ \label{alg:fuzz:mutation_loop}
                $skeleton \leftarrow \FnSkeletonize{f}$ \label{alg:fuzz:skeletonize}
                \BlankLine
                $g \leftarrow \FnRandomSelect{G}$ \label{alg:fuzz:select_generator}\; 
                $terms \leftarrow \FnGenerate{g, skeleton}$ \label{alg:fuzz:generate_terms}\; 
                $f \leftarrow \FnSynthesize{terms}$ \label{alg:fuzz:synthesize}\; 
                \BlankLine
                $Results \leftarrow \FnValidate{f, Solvers}$\; \label{alg:fuzz:validate}
                \If{\FnDiscrepancy{$Results$}}{ \label{alg:fuzz:discrepancy}
                    $Bugs \leftarrow Bugs \cup \{f\}$ \label{alg:fuzz:add_bug}
                }
            }
        }
    }
    \caption{\small Main fuzzing loop of \textbf{\tool}.}
    \label{alg:fuzz}
\end{algorithm}

\subsection{Skeleton-guided Mutation}

\tool~generates diverse test formulas using a \emph{skeleton-guided mutation} strategy (Algorithm~\ref{alg:fuzz}).
This design choice enables \tool~to explore deeper solver behaviors while overcoming gaps in the SMT-LIB specification.
For instance, certain constructs such as the quantifiers \texttt{forall} and \texttt{exists} are not fully covered by SMT-LIB theory documentation, limiting what a purely grammar-based generator can produce. 
By leveraging existing formulas as structural templates, \tool~extends the expressiveness of generated inputs and reaches behaviors that grammar-only methods cannot.
Specifically, the process begins with two inputs: a collection of seed formulas, $Seeds$, and a set of term generators, $G$, obtained from the previous phase. 
These are combined with lightweight structural synthesis to create a wide variety of formulas that are effective at uncovering solver bugs.
In each iteration, \tool~first randomly selects a seed formula $f$ from $Seeds$ (Line~\ref{alg:fuzz:select_seed}).
\tool~then derives the \emph{skeleton} of the formula by randomly removing a subset of its atomic formulas, \ie~basic sub-formulas without logical connectives like \texttt{and} and \texttt{or}, and replacing them with a \colorbox{gray!40!white}{\texttt{<placeholder>}} (Line~\ref{alg:fuzz:skeletonize}). 
This transformation enables flexible insertion of Boolean terms while preserving the overall structure and validity of the formula.
Next, \tool~selects a generator $g \in G$ (Line~\ref{alg:fuzz:select_generator}) at random and uses it to produce a set of Boolean terms (Line~\ref{alg:fuzz:generate_terms}). 
Before inserting these terms into the skeleton, \tool~checks for matching sorts (\eg, \texttt{Bool}, \texttt{Int}, or \texttt{Real}) between the variables in the generated terms and those in the skeleton. 
When compatible, variables in the generated terms are randomly replaced with corresponding variables from the skeleton to increase semantic interaction between the inserted content and the original structure.

These modified terms are then integrated into the skeleton, yielding a new formula $f$ (Line~\ref{alg:fuzz:synthesize}). 
The formula is then validated using multiple solver implementations (Line~\ref{alg:fuzz:validate}).
If the solvers produce inconsistent results or one crashes, the formula is recorded in the bug list $Bugs$ (Lines~\ref{alg:fuzz:discrepancy} - \ref{alg:fuzz:add_bug}). 
In cases where the formula includes solver-specific features, we compare results across different versions of the same solver. 
To identify the source of a discrepancy, we use the \texttt{get-model} command to extract a model, \ie~a set of variable assignments, from the solver that returns \texttt{sat}. 
We then re-evaluate the formula against all solvers using this model. 
If the formula definitely evaluates to \texttt{sat} with the model, we report a soundness bug in any solver that returned \texttt{unsat}, and vice versa. 
If the model fails to satisfy the formula, we also classify it as an invalid model bug.
This mutation-validation process continues until a user-defined stopping condition is met, steadily generating new formulas for bug hunting.

\begin{figure*}[t]
\centering
\begin{minipage}[t]{0.43\linewidth}
\centering
\begin{subfigure}[t]{\linewidth}
\begin{lstlisting}[basicstyle=\footnotesize\ttfamily]
(declare-fun T () Int)
(assert (or (= T 0) (< T 1)))
(check-sat)
\end{lstlisting}
\caption{Original seed formula.}
\label{ex:seed}
\end{subfigure}

\vspace{1em}

\setcounter{subfigure}{2}
\begin{subfigure}[t]{\linewidth}
\vspace{0.5em}
\begin{lstlisting}[basicstyle=\footnotesize\ttfamily]
((_ divisible 3) (mod int0 3))
(= str0 "")
\end{lstlisting}
\caption{Boolean terms generated from generators.}
\label{ex:terms}
\end{subfigure}
\end{minipage}
\hfill
\begin{minipage}[t]{0.53\linewidth}
\centering
\setcounter{subfigure}{1}
\begin{subfigure}[t]{\linewidth}
\begin{lstlisting}[basicstyle=\footnotesize\ttfamily]
(declare-fun T () Int)
(assert (or <@\colorbox{gray!40}{<placeholder>}@> <@\colorbox{gray!40}{<placeholder>}@>))
\end{lstlisting}
\caption{Skeleton extracted from the seed.}
\label{ex:skeleton}
\end{subfigure}

\vspace{1em}

\setcounter{subfigure}{3}
\begin{subfigure}[t]{\linewidth}
\vspace{-0.7em}
\begin{lstlisting}[basicstyle=\footnotesize\ttfamily]
(declare-fun T () Int)
(declare-fun str0 () String)
(assert (or ((_ divisible 3) (mod T 3)) (= str0 "")))
(check-sat)
\end{lstlisting}
\caption{Synthesized formula.}
\label{ex:final}
\end{subfigure}
\end{minipage}

\caption{Illustrative process of \tool's skeleton-guided mutation.}
\label{fig:illustrative}
\end{figure*}

\mypara{Illustrative Example in Steps.}
We illustrate the skeleton-guided mutation process with a concrete example, shown in Figure~\ref{fig:illustrative}. The procedure unfolds in three steps:

\begin{itemize}[label={},leftmargin=1em,itemindent=-1em]

\item \textbf{Step 1: Skeleton extraction.} 
\tool~first selects a seed formula from the seed pool (Figure~\ref{ex:seed}). 
It then abstracts away certain atomic formulas at random, replacing them with \colorbox{gray!40}{\texttt{<placeholder>}} markers, as depicted in Figure~\ref{ex:skeleton}. 
This preserves the original logical structure while leaving positions open for new terms.

\item \textbf{Step 2: Generator invocation.} 
\tool~then invokes generators from $G$ to produce candidate Boolean terms. 
Notably, generators from different theories can be used simultaneously. 
In this example, two terms are produced by the \texttt{Int} and \texttt{String} generators (Figure~\ref{ex:terms}), respectively. 
Before inserting them into the skeleton, \tool~checks sort compatibility with variables from the original seed and adapts variables as needed. 
For instance, the variable \texttt{int0} in the first term depicted in Figure~\ref{ex:terms} is replaced with the existing integer variable \texttt{T} from the seed formula, enhancing semantic interactions in mutants. 
This adaptation is especially valuable for complex seed formulas, where only a subset of sub-formulas may be replaced.

\item \textbf{Step 3: Formula synthesis.} 
Finally, the adapted terms are inserted into the skeleton to produce a new formula (Figure~\ref{ex:final}). 
This synthesized formula is then tested across multiple solvers.  
Any crash or inconsistency in solver outputs indicates a potential bug.

\end{itemize}

This example is adapted from a real crash in \href{https://github.com/cvc5/cvc5/issues/12058}{cvc5\#12058} detected by \tool, which was later fixed by developers.

\subsection{Implementation}

We implemented \tool~as a practical and extensible fuzzing tool for SMT solvers using Python.
Specifically, for the LLM-assisted generator construction phase, we utilize GPT-4~\cite{DBLP:journals/corr/abs-2303-08774} to assist in summarizing SMT grammars, implementing generators, and refining them through self-correction.
We selected GPT-4 due to its strong performance on programming tasks and its ability to follow natural language instructions effectively~\cite{DBLP:journals/corr/abs-2303-08774}.
In the skeleton-guided mutation phase, \tool~applies ten iterations of mutation to each input seed formula, a configuration commonly employed in relevant research practices~\cite{DBLP:journals/pacmpl/ParkWZS21}.

\section{Experimental Evaluation}
\label{sec:evaluation}
\newcommand{\cis}{C}
\newcommand{\nis}{\overline{C}}

This section presents a comprehensive evaluation of the effectiveness of \tool.

\subsection{Evaluation Setup}
\label{subsec:EvaluationSetup}

\mypara{Research Questions.}
The experiments conducted aim to answer the following research questions:

\begin{itemize}[leftmargin=*]
\item \textbf{RQ1:} Can \tool~be used to expose new real bugs in SMT solvers? (Section~\ref{sec:rq1})
\item \textbf{RQ2:} How does \tool~compare with state-of-the-art SMT solver fuzzers? (Section~\ref{sec:rq2})
\item \textbf{RQ3:} How do \tool's different components contribute to its overall performance? (Section~\ref{sec:rq3})

\end{itemize}

\mypara{Types of Bugs.}
Following the practices of previous studies~\cite{winterer2020validating,DBLP:journals/pacmpl/WintererZS20,yao2021sae,sunvalidating}, we categorize the bugs found in SMT solvers into three main types.
The three categories of bugs are defined as follows:

\noindent $\bullet$ \textit{Soundness bugs}: This type of bug occurs when two solvers provide opposite results for the same formula, where one solver reports \texttt{sat} while the other reports \texttt{unsat}. Such inconsistencies are considered soundness bugs.

\noindent $\bullet$ \textit{Invalid model bugs}: An invalid model bug occurs when a solver correctly identifies a formula as satisfiable (\texttt{sat}), but the provided model fails to satisfy the constraints in the formula.

\noindent $\bullet$ \textit{Crash bugs}: This type of bug occurs when a solver exhibits abnormal behavior during the solving process, such as assertion violations and segmentation faults.

\mypara{Environment.}
We conduct all of our experiments on a machine equipped with an Intel Xeon CPU Gold-6248 processor (20 cores and 256GB RAM)  running a Docker container with the Ubuntu 20.04 64-bit operating system. 
In line with prior studies~\cite{yao2021sae,sunvalidating}, we set the time limit for solving queries for each test formula to 10 seconds.

\mypara{Test Seeds.}
\tool~is a general-purpose fuzzer that accepts any existing SMT formula as input.
Over the years, the SMT community has accumulated a vast collection of formulas for testing and benchmarking.
For instance, in addition to the SMT-LIB official benchmark suite~\cite{SMT-COMP}, prior work~\cite{sunvalidating,sunsmt} curated approximately 3,700 bug-triggering formulas from issue trackers of major solvers including Z3 and cvc5, and found them to be of higher quality in solver testing.
In our experiments, we use the bug-triggering formulas released by these studies~\cite{sunvalidating,sunsmt} as seed inputs, as they demonstrate high bug-finding potential while keeping the seed set size manageable for fair comparison.
To avoid data leakage, we re-execute all seed formulas on the latest solver versions and remove any that still trigger previously reported bugs.

\subsection{RQ1: Bug Detection}
\label{sec:rq1}

The objective of \textit{RQ1} is to examine whether \tool~can detect new real bugs in SMT solvers.

\subsubsection{\textbf{Experimental Setup}}
\

\mypara{Targeted Solvers.} 
We evaluate our approach on Z3 and cvc5, two widely used SMT solvers that support a broad range of SMT-LIB features and have been extensively studied in prior work~\cite{DBLP:journals/pacmpl/WintererZS20,DBLP:journals/pacmpl/ParkWZS21,sunvalidating}. 
Following common practice~\cite{DBLP:journals/pacmpl/ParkWZS21,sunvalidating}, we primarily test solvers in their default configuration, which reflects standard usage by practitioners.
We enable a small set of essential options for detecting invalid models, including \texttt{model\_validate=true} in Z3 and \texttt{--check-models} in cvc5, which are generally considered part of the default mode~\cite{DBLP:journals/pacmpl/WintererZS20}. 
We additionally explore selected non-default options recommended by solver developers, such as the cvc5 fuzzing guidelines~\cite{cvc5guidelines}, in our bug-hunting campaign. 
All experiments are conducted on the latest trunk versions to avoid reporting duplicate bugs.

\mypara{Bug Inspection and Reduction.\label{sec:reduction}}
We follow a semi-automated workflow to inspect and reduce potential bugs detected by \tool. Our goal is to avoid redundant reports, minimize manual effort, and provide developers with clear, minimal reproducing cases.
For crashes, \tool\ automatically clusters failing cases using their crash stacks. We treat all crashes that reach the same code location as a single issue. This grouping requires minimal manual inspection. Exception information reported by solvers is used to further refine de-duplication.
For soundness and invalid model issues, we group cases by the theory involved, following prior work~\cite{winterer2020validating}. A lightweight script identifies the theory of each formula, enabling us to manually inspect only one representative per group. Once developers provide a fix, all remaining cases in that group are automatically revalidated.
Furthermore, to improve developer understanding and ease the patching process, we reduce bug-triggering formulas using delta-debugging tools. We primarily rely on \texttt{ddSMT}~\cite{DBLP:conf/cav/KremerNP20}. 
For cases not supported by \texttt{ddSMT}, such as those involving certain solver-specific theories, we instead apply \textsc{C-Reduce}~\cite{DBLP:conf/pldi/RegehrCCEEY12}.
The resulting minimized formulas help ensure concise and actionable reports.

\subsubsection{\textbf{Results}}
\
\begin{figure}[t]
    \centering
    \begin{minipage}[b]{0.43\textwidth}
        \centering
        \captionof{table}{Status of bugs found in the solvers.}
        \vspace{1em}
        \bugcounttable
        \label{fig:bugcount-table}
    \end{minipage}
    
    \begin{minipage}[b]{0.43\textwidth}
        \centering
        \captionof{table}{Bug types among the reported bugs.\label{fig:bugtype-table}}
        \vspace{1em}
        \bugtypetable
      \end{minipage}

\end{figure}

\mypara{Statistics of Bugs.}
\tool~generated a large and diverse input set during our bug-hunting campaigns. 
Across all runs, it produced approximately ten million test cases. 
On average, each formula is 4{,}828 bytes, and using these inputs, \tool~identified 727 bug-triggering formulas, which were then triaged and reported to the developers of the respective solvers.
Table~\ref{fig:bugcount-table} presents the status of the unique bugs discovered by \tool~in Z3 and cvc5.
In total, \tool~has reported \reported~bugs across the three solvers, of which \confirmed~bugs have been confirmed by the developers.
As of the time of writing, \fixed~bugs have been fixed, leaving the remaining bugs awaiting processing by the developers.
Despite our best efforts to de-duplicate the bugs before reporting them, it is impossible to guarantee the complete absence of duplication.
However, the small number of duplicate bugs suggests that the de-duplication process is effective.
Notably, none of the bugs reported by \tool~have been marked as \textit{won't fix} or \textit{invalid} by the developers, indicating that the bugs identified by \tool~are valid and that our pre-reporting inspection process is effective.

Table~\ref{fig:bugtype-table} depicts the distribution of the bug types of the confirmed bugs.
Specifically, the majority of the confirmed bugs (35 out of \confirmed) are crash bugs.
The remaining bugs, including six invalid model bugs and four soundness bugs, affect the correctness of the solvers.
These results demonstrate that \tool~is capable of exposing real bugs in SMT solvers.
Note that among the bugs discovered by \tool, \textbf{11 bugs involve newly added or solver-specific theories} that existing SMT solver fuzzers are fundamentally incapable of uncovering, either by design or in principle.
We present concrete examples of these bugs in Section~\ref{subsec:BugSample}.

\begin{figure}[t]
    \centering
    \begin{subfigure}{0.49\textwidth}
        \centering
        \releaseZ
    \caption{Z3}

    \end{subfigure}

\vspace{1em}

    \begin{subfigure}{0.49\textwidth}
        \centering
        \releaseCVC
    \caption{cvc5}
    \end{subfigure}
\caption{Confirmed bugs that affect the corresponding release versions of Z3 and cvc5. \label{fig:affect}}
\end{figure}

\mypara{Bug Lifespan.}
To further understand the bugs unveiled by \tool, we conducted an analysis of their impact on the release versions of Z3 and cvc5, thereby revealing the bugs' lifespans. 
To investigate the longevity of bugs, similar to prior work~\cite{sunsmt}, we select Z3 release versions from 4.8.1 to 4.13.0 and cvc5 release versions from 0.0.2 to 1.2.0 as the subjects.
Notably, Z3-4.8.1 was released over five years ago, in October 2018, while cvc5-0.0.2  the first official release of cvc5 in October 2021. 
We evaluated each solver's release version against the test instances of confirmed bugs reported by \tool~for Z3 and cvc5, determining the persistence of these bugs across versions. A bug is considered to affect a version if the original formula successfully triggers the bug.
Figure~\ref{fig:affect} illustrates the distribution of confirmed bugs affecting the solvers' release versions.
Our analysis reveals that most of the bugs were only present in the trunk versions of the solvers.
Specifically, three of the bugs in Z3 remained latent for over six years, indicating that these bugs were not exposed by the developers or other SMT testing tools.
In summary, the results suggest that \tool~is capable of identifying long-standing bugs in SMT solvers but is particularly effective in uncovering newly introduced bugs that may result from the ongoing development of solvers. This underscores the practical significance of our approach.

\mypara{Developers' Feedback.}
Several issues reported by \tool\ have sparked active discussions among solver developers and received positive recognition. 
For instance, one developer gratefully acknowledged, ``\textit{Thank you for this report! This was a new bug that I introduced in my last patch (whoops!)}'' 
Such comments highlight the immediate impact of our findings. Another developer commendation, stating, ``\textit{Credits to the reporter for finding the problem},'' further underscores the value developers place on these reports.
Besides, some filed issues were labeled as \textit{major}\footnote{\url{https://github.com/cvc5/cvc5/pull/10150}}, reflecting its critical nature. 

\begin{findingbox}
\textit{\tool~is effective in detecting real bugs in SMT solvers, including long-latent bugs as well as bugs on newly added and solver-specific theories.}
\end{findingbox}

\subsection{RQ2: Comparison with Existing Fuzzers}
\label{sec:rq2}

To comprehensively evaluate the effectiveness of \tool, we compare it with the state-of-the-art fuzzers, including STORM, YinYang, OpFuzz, TypeFuzz, HistFuzz, and Fuzz4All.
They are all open-source fuzzers proposed in recent years, capable of supporting multiple logics and identifying a significant number of bugs within SMT solvers.
We use the latest versions of these fuzzers provided by their developers in our experiments. Regarding the targeted solvers, we utilize the latest release versions of Z3 (4.14.0) and cvc5 (1.2.1).
In this RQ, we investigate the achieved code coverage and the number of unique bugs detected by different fuzzers.

\begin{figure*}
    \centering
    \begin{subfigure}[b]{0.4\textwidth}
        \centering
        \includegraphics[width=\linewidth]{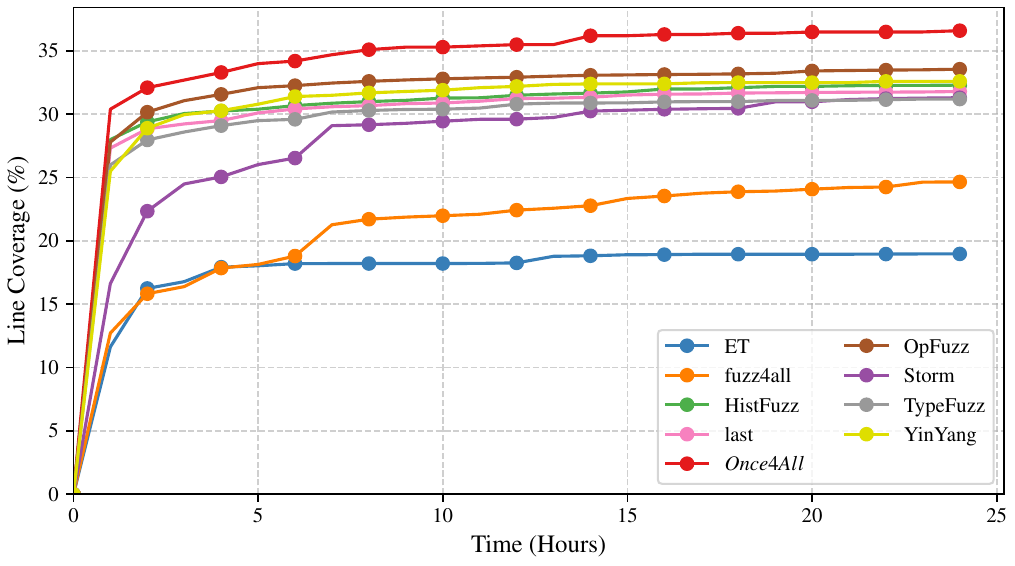}
        \vspace{-1em}
        \caption{Line coverage on Z3.}
        \label{fig:coverage-z3-line}
    \end{subfigure}
    \hspace{2em}
    \begin{subfigure}[b]{0.4\textwidth}
        \centering
        \includegraphics[width=\linewidth]{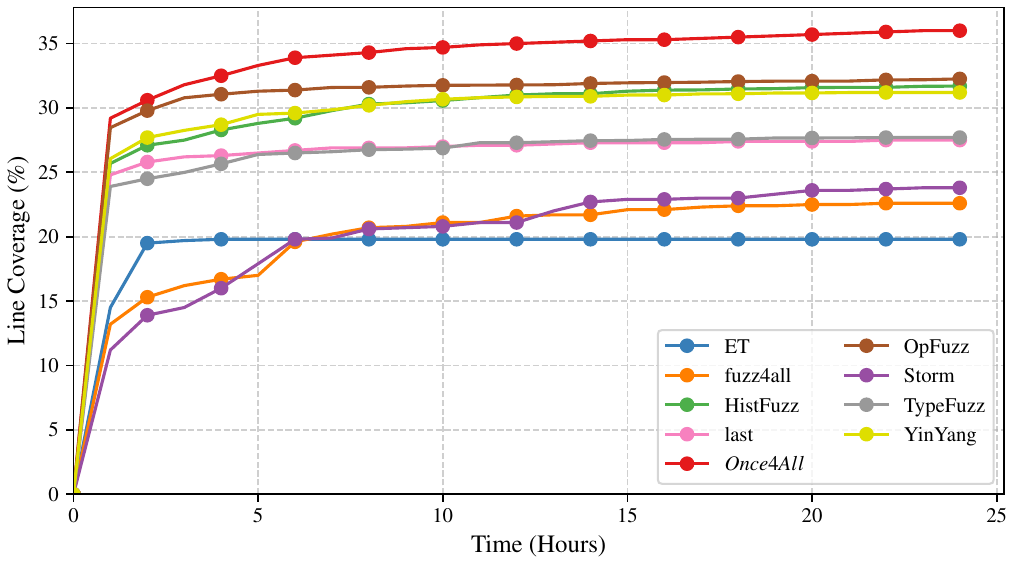}
        \vspace{-1em}
        \caption{Line coverage on cvc5.}
        \label{fig:coverage-cvc5-line}
    \end{subfigure}

    \begin{subfigure}[b]{0.4\textwidth}
        \centering
        \includegraphics[width=\linewidth]{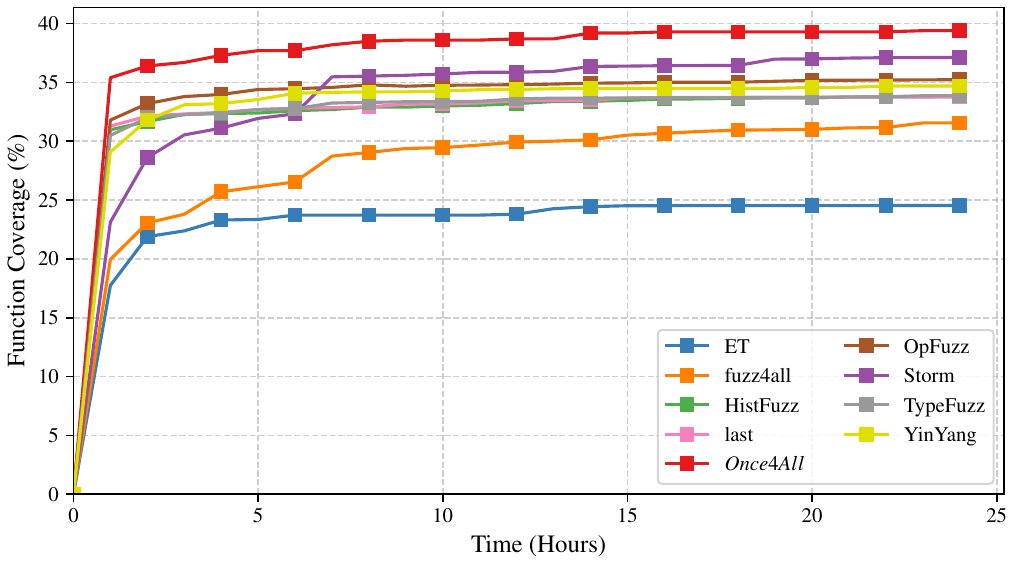}
        \vspace{-1em}
        \caption{Function coverage on Z3.}
        \label{fig:coverage-z3-function}
        
    \end{subfigure}
    \hspace{2em}
    \begin{subfigure}[b]{0.4\textwidth}
        \centering
        \includegraphics[width=\linewidth]{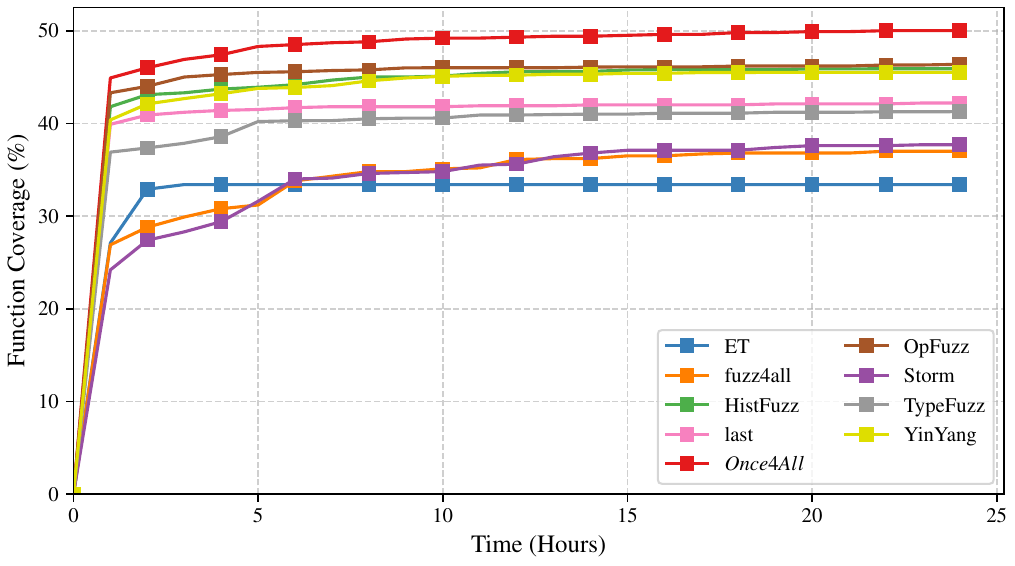}
        \vspace{-1em}
        \caption{Function coverage on cvc5.}
        \label{fig:coverage-cvc5-function}
    \end{subfigure}
    \vspace{-0.5em}
    \caption{Code coverage growth over time for different fuzzers on Z3 and cvc5.}
    \label{fig:coverage-growth} 
\end{figure*}

\mypara{Code Coverage.}
To evaluate the effectiveness of \tool\ in exploring SMT solver implementations, we measure the code coverage it achieves in comparison with baseline fuzzers on Z3 and cvc5.
Each fuzzer is executed for 24 hours, and coverage growth is recorded at one-hour intervals.
This experimental duration and measurement setup follow established practice in prior work~\cite{10.1145/3622781.3674171}. 
To ensure a fair comparison, all fuzzers are initialized with the same seed inputs used in \textit{RQ1}.
Likewise, both solvers and fuzzers are executed in their default configurations, without enabling optional features or performing parameter tuning.
We measure code coverage using \texttt{gcov}\footnote{\url{https://gcc.gnu.org/onlinedocs/gcc/Gcov.html}}, which tracks the executed source lines and functions in each solver during fuzzing. 
Figure~\ref{fig:coverage-growth} present the growth of line and function coverage on Z3 and cvc5, respectively, over the 24-hour period.
Overall, \tool~consistently outperforms the baselines on both solvers. 
Across both solvers, all fuzzers show a steady increase in coverage over time and approach a saturation point after approximately 20 hours.
Throughout the entire execution, however, \tool\ consistently achieves higher coverage than the baseline approaches at each measurement interval.
Overall coverage remains below 50\% in most cases, with only function coverage on cvc5 slightly exceeding this threshold.
This result is expected under our experimental design and consistent with prior findings in testing complex system software.
Similar to compiler testing, achieving high absolute code coverage in SMT solvers is inherently difficult: even state-of-the-art compiler testing techniques report line coverage below 40\%~\cite{DBLP:journals/pacmpl/LiTS24}.
Moreover, using default solver configurations to reflect realistic usage leaves substantial functionality disabled and large portions of the codebase unreachable by design.
Consequently, code coverage for SMT solvers is generally lower than that reported for fuzzers targeting applications or libraries, where coverage can exceed 60\%~\cite{DBLP:conf/ndss/WangJLZBWS20}.

To further examine the differences among fuzzers, we manually analyze the final coverage results after 24 hours.
We find that \tool\ is complementary to existing fuzzers rather than fully subsuming them.
Specifically, while it does not cover all code regions reached by other fuzzers, \tool\ exercises additional parts of the solvers that none of the baseline fuzzers reach.
For example, on cvc5, \tool\ covers solver-specific theory implementations such as the files under the \texttt{src/theory/sets/} directory, which implement the theory of sets.
However, these files are not reached by any of the baseline fuzzers.
In summary, these results show that \tool\ achieves higher code coverage than existing SMT solver fuzzing techniques.
By exercising a broader portion of the solvers' codebases, \tool\ enables more comprehensive exploration of solver behavior.

\begin{figure}
    \centering
    \includegraphics[width=0.36\textwidth]{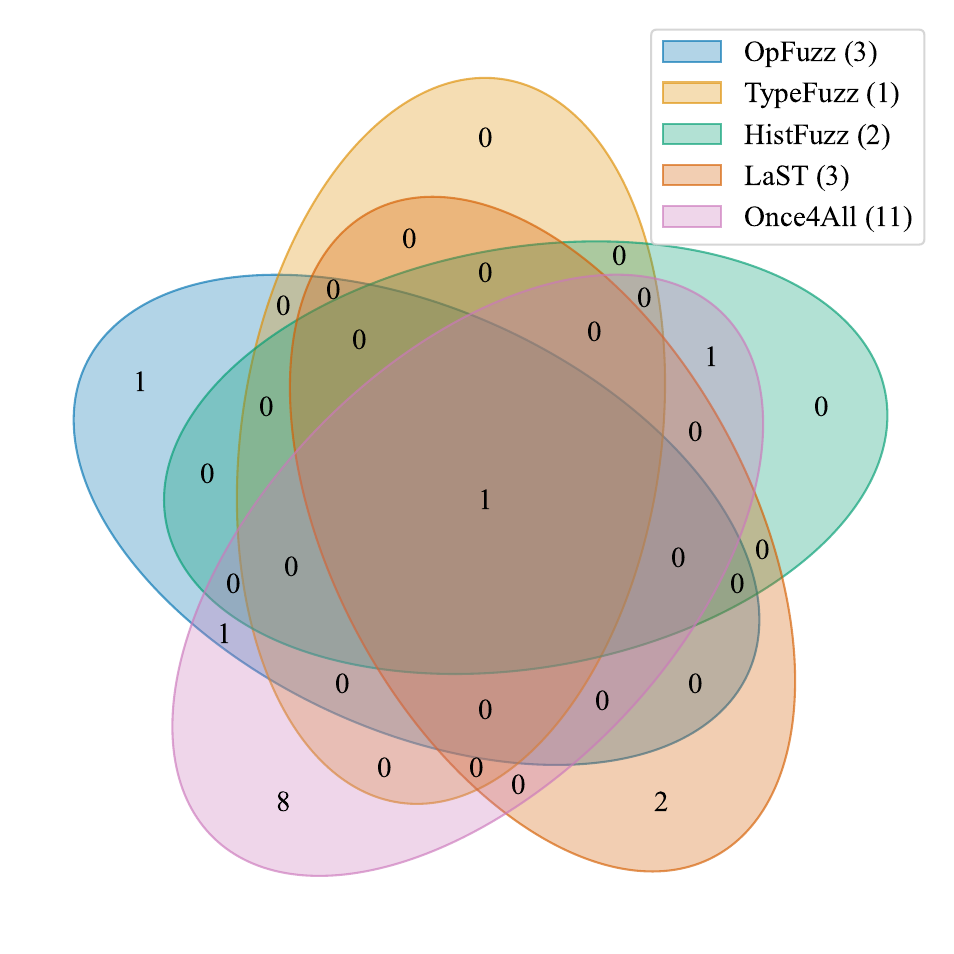}
    \vspace{-1em}
    \caption{The distribution of unique known bugs detected by \tool~and the baselines on previous versions of solvers.}
    \label{fig:previous}
\end{figure}

\mypara{Bug-Finding Capability.}
We run \tool\ for 24 hours and record the number of unique known bugs found by each fuzzer, consistent with prior work~\cite{sunsmt}.
The seeds employed and the baseline techniques mirror those utilized in the prior code coverage experiments.
To identify the unique bugs, we employed the \textit{Correcting Commit} approach~\cite{DBLP:conf/icse/ChenHHXZ0X16,sunvalidating}. 
Specifically, we feed a potential bug-triggering instance generated by a fuzzer to different commit versions of the solvers and check whether the bug was fixed in one commit. 
If a bug can be triggered before a commit while cannot be triggered after the commit, we consider this commit as the correcting commit of the bug.
The bugs corresponding to different correcting commits are considered as different bugs.
In practice, we exploit binary search to accelerate the process of identifying correcting commits.
Note that we focus on using fuzzing tools to find known bugs that have already been resolved, rather than finding new bugs in this experiment, which enables us to conveniently determine the number of unique bugs.
The findings, presented in Figure~\ref{fig:previous}, reveal that \tool~excels in detecting the highest number of unique bugs across the solvers. 
Previous techniques are incapable of identifying more than three bugs, leading us to infer that these methods encounter difficulties in detecting newly introduced bugs in recent solver versions.

In conclusion, the experiments highlight \tool's advantage in both bug detection and code coverage efficacy.

\begin{findingbox}
\textit{\tool~outperforms the state-of-the-art fuzzing techniques in terms of bug-finding capability and code coverage. Specifically, \tool~achieves better performance on cvc5, which is supposed to stem from the fact that cvc5 has more extended theories.}
\end{findingbox}

\begin{figure*}
    \centering
    \begin{subfigure}[b]{0.4\textwidth}
        \centering
        \includegraphics[width=\linewidth]{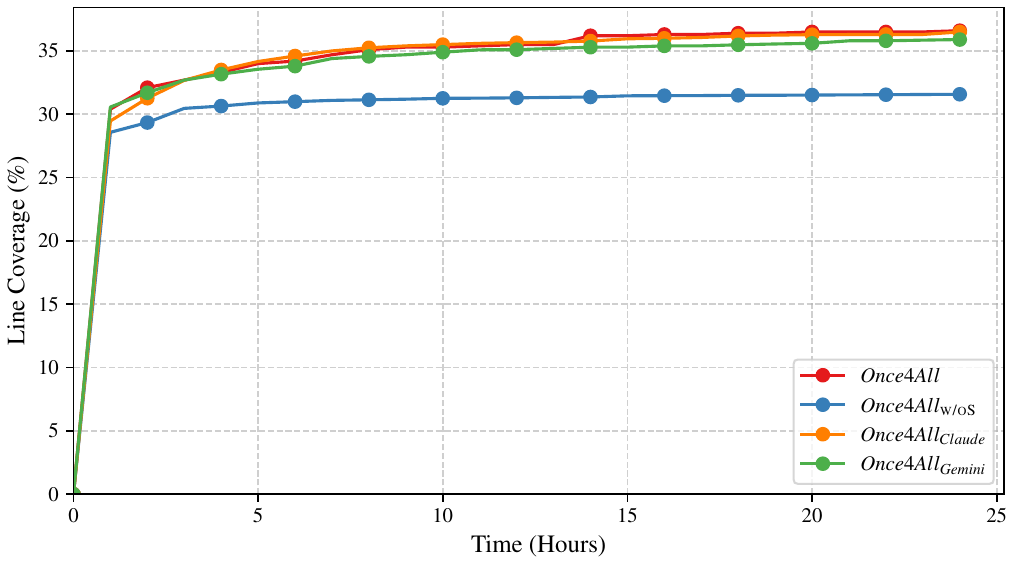}
        \vspace{-1em}
        \caption{Line coverage on Z3.}
        \label{fig:coverage-ablation-z3-line}
    \end{subfigure}
    \hspace{2em}
    \begin{subfigure}[b]{0.4\textwidth}
        \centering
        \includegraphics[width=\linewidth]{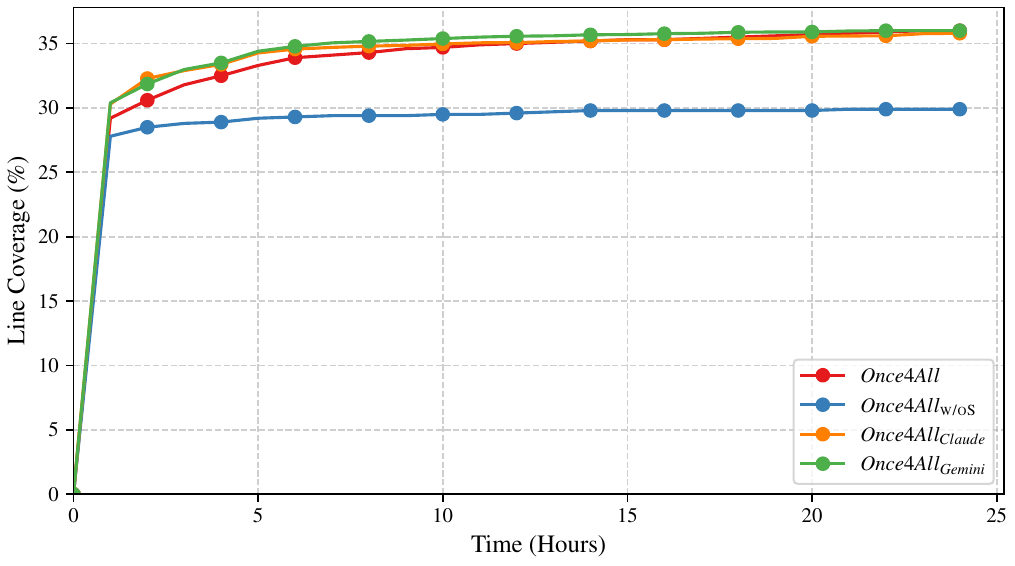}
        \vspace{-1em}
        \caption{Line coverage on cvc5.}
        \label{fig:coverage-ablation-cvc5-line}
    \end{subfigure}

    \begin{subfigure}[b]{0.4\textwidth}
        \centering
        \includegraphics[width=\linewidth]{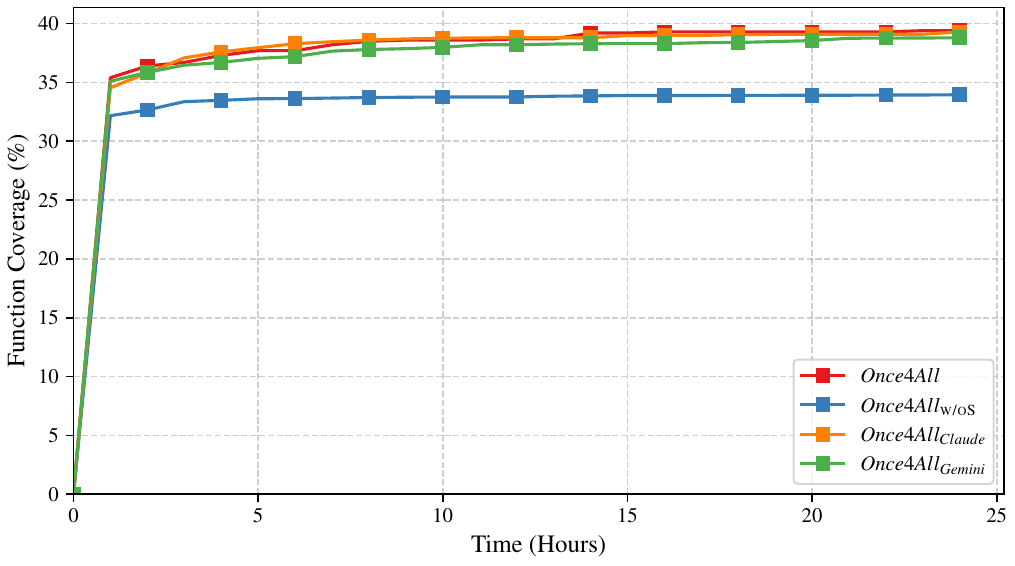}
        \vspace{-1em}
        \caption{Function coverage on Z3.}
        \label{fig:coverage-ablation-z3-function}
        
    \end{subfigure}
    \hspace{2em}
    \begin{subfigure}[b]{0.4\textwidth}
        \centering
        \includegraphics[width=\linewidth]{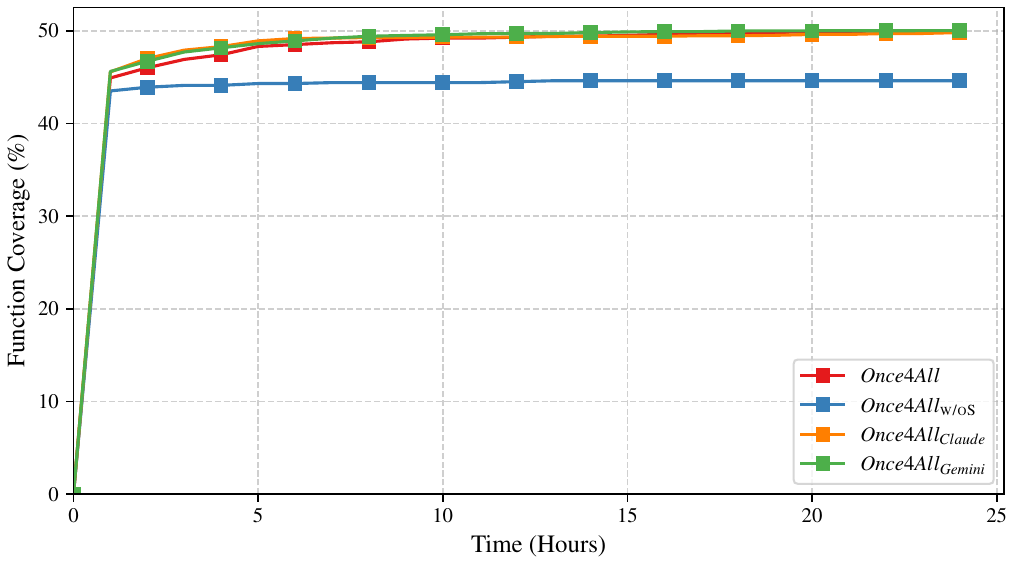}
        \vspace{-1em}
        \caption{Function coverage on cvc5.}
        \label{fig:coverage-ablation-cvc5-function}
    \end{subfigure}
    \vspace{-0.5em}
    \caption{Code coverage growth over time for \tool\ and its variants on Z3 and cvc5.}
    \label{fig:coverage-ablation}
\end{figure*}

\subsection{RQ3: Sensitivity Analysis}
\label{sec:rq3}

This research question examines how individual design choices in \tool\ affect its effectiveness. In particular, we focus on two factors: the role of skeleton-guided synthesis and the choice of the underlying LLM.
Specifically, to quantify the contribution of skeleton-guided synthesis, we construct a variant of \tool, denoted as $\tool_{\textsc{w/oS}}$ (\emph{without Skeletons}). 
This variant generates test formulas solely using LLM-synthesized generators and does not incorporate skeletons extracted from existing formulas. 
By comparing \tool\ with $\tool_{\textsc{w/oS}}$, we isolate the impact of skeleton guidance and establish a baseline that reflects the effectiveness of the LLM component alone.

In addition to the synthesis strategy, we investigate whether \tool's performance is sensitive to the choice of LLM. To this end, we create two additional variants of \tool, each using a different LLM for grammar summarization and generator synthesis:
\begin{itemize}[leftmargin=*, ]
  \item $\tool_{Gemini}$, which uses  Gemini 2.5 Pro~\cite{DBLP:journals/corr/abs-2507-06261};
  \item $\tool_{Claude}$, which uses  Claude 4.5 sonnet~\cite{anthropic2025claude}.
\end{itemize}

\begin{figure}
    \centering
    \includegraphics[width=0.33\textwidth]{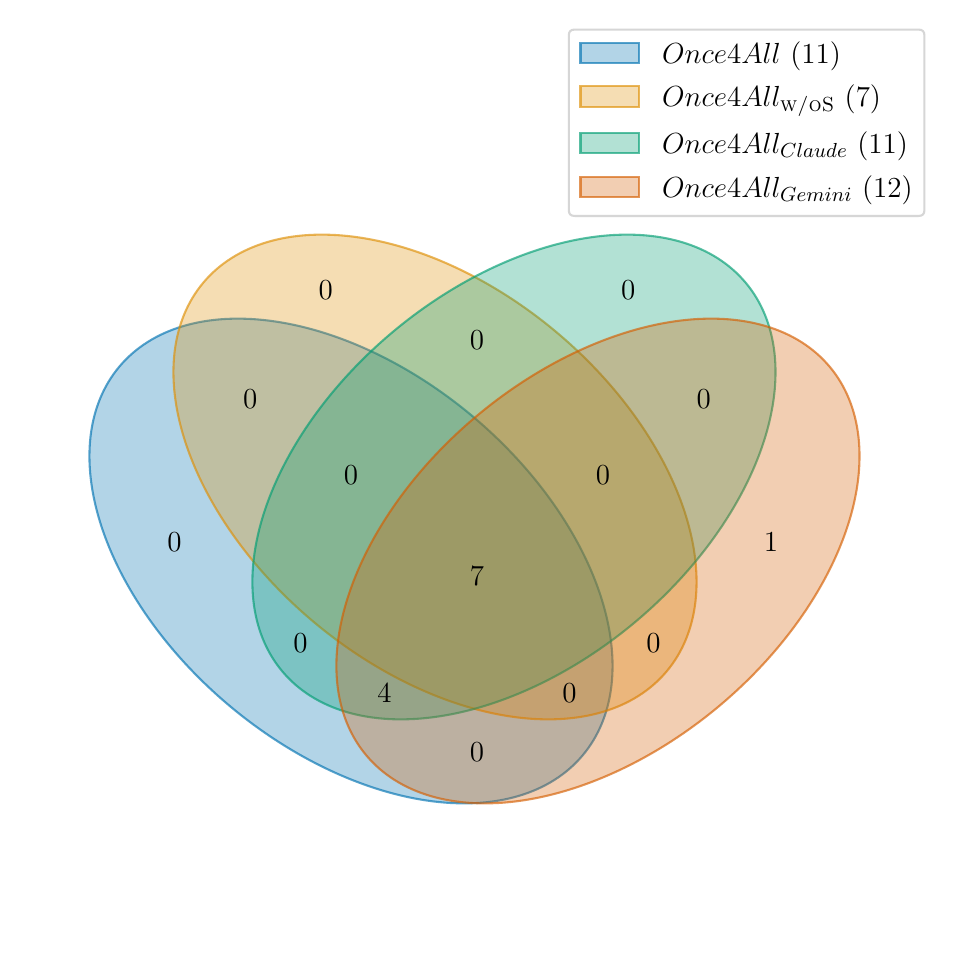}
    \vspace{-2em}
    \caption{The distribution of unique known bugs detected by \tool~and its variants on Z3 and cvc5.}
    \label{fig:bug-ablation}
\end{figure}

All variants are evaluated under the same experimental setup as in \textit{RQ2}. 
We measure both code coverage and bug-finding effectiveness to assess how these design choices influence \tool's overall performance.
Figure~\ref{fig:coverage-ablation} shows the growth of code coverage over time for \tool\ and its variants on Z3 and cvc5.
Removing skeleton guidance leads to a clear degradation in coverage: $\tool_{\textsc{w/oS}}$ consistently explores fewer code paths on both solvers.
In contrast, the variants using alternative LLMs, $\tool_{Gemini}$ and $\tool_{Claude}$, achieve coverage comparable to the original \tool, with only minor differences.
The reduced coverage of $\tool_{\textsc{w/oS}}$ also translates into weaker bug-finding performance.
As shown in Figure~\ref{fig:bug-ablation}, this variant detects only a subset of the bugs uncovered by \tool.
Although $\tool_{\textsc{w/oS}}$ remains effective to some extent, skeleton guidance substantially improves the ability to exercise deeper solver behaviors and expose additional bugs.
Meanwhile, both $\tool_{Gemini}$ and $\tool_{Claude}$ identify a similar number of bugs as \tool, indicating that the framework generalizes well across different LLMs.

\begin{findingbox}
\textit{Skeleton guidance plays a key role in \tool's effectiveness, significantly improving both code coverage and bug discovery. Using different LLMs yields comparable results, demonstrating that \tool's design is robust to the choice of LLM.}
\end{findingbox}

\begin{figure*}

\begin{minipage}{0.46\textwidth}
\begin{subfigure}{\textwidth}
\centering
\begin{lstlisting}[basicstyle=\footnotesize\ttfamily]
(set-logic QF_FF)
(declare-const v (_ FiniteField 3))
(assert (= v (ff.bitsum (ff.mul v v) 
        (as ff-1 (_ FiniteField 3)))))
(check-sat)
\end{lstlisting}
\caption{\normalsize Invalid model issue regarding Finite Field theory in cvc5. \url{https://github.com/cvc5/cvc5/issues/11969}
\label{sample:a}}
\vspace{0.3cm}
\end{subfigure}

\begin{subfigure}{\textwidth}
\centering
\begin{lstlisting}[basicstyle=\footnotesize\ttfamily]
(declare-fun s () (Set UnitTuple))
(assert (rel.join s (as set.empty (Set UnitTuple))))
\end{lstlisting}
\caption{\normalsize Segmentation fault issue regarding Set theory in cvc5. \url{https://github.com/cvc5/cvc5/issues/11903}
\label{sample:b}}
\vspace{0.3cm}
\end{subfigure}
\end{minipage}
\hfill
\begin{minipage}{0.46\textwidth}
\begin{subfigure}{\textwidth}
\centering
\begin{lstlisting}[basicstyle=\footnotesize\ttfamily]
(declare-const x15 Bool)
(declare-const x Real)
(declare-const x1 Real)
(declare-const x9 Bool)
(declare-fun v () Real)
(assert (forall ((r Real)) (or x9 
(or (= (+ r 1.0) (mod 0 (to_int x)))))))
(assert (and (> 0.0 x1) 
(< x (/ 1.0 (* v x))) (<= 0.0 (/ 0 v))))
(check-sat)
\end{lstlisting}
\caption{\normalsize Segmentation fault issue in terms of standard theory in Z3. \url{https://github.com/Z3Prover/z3/issues/6935}}
\label{sample:c}
\vspace{0.3cm}
\end{subfigure}

\end{minipage}

\vspace{0.5em}

\caption{Selected bug samples in Z3 and cvc5. \label{fig:bug-array-samples}}
\end{figure*}

\subsection{Bug Sample}
\label{subsec:BugSample}
To demonstrate \tool's effectiveness in uncovering bugs across both standard and newly extended theories, 
we present several representative case studies.

Figure~\ref{sample:a} shows a bug in cvc5's finite field theory, which is an extended theory introduced in 2022.
In this case, the \texttt{ff.bitsum} operator incorrectly ignored coefficient multipliers for constant children. 
As explained by the developer, the constraint encodes the equation $v = v^2 + 2 \mod 3$ using \texttt{ff.bitsum}, expecting solutions $v = 1$ and $v = 2$ in $\mathbb{F}_3$. 
However, due to a faulty implementation that added constants without scaling them, the solver misinterpreted the formula as $v = v^2 + 1$, leading to wrong models. 
The root cause was a missing multiplication by the coefficient in the accumulation logic. 
The patch fixed this by ensuring proper weighting of constant terms. This case highlights how errors can silently compromise solver correctness, especially in extended theories that are not sufficiently exercised.

Figure~\ref{sample:b} exposes a bug in cvc5's set theory, which is also an extended theory specifically customized in cvc5.
In this case, the solver failed to reject invalid join operations over nullary relations—sets of \texttt{UnitTuple}, which have arity zero. The SMT input attempts to join two such sets, expecting a type error because relational joins are undefined for nullary inputs. Instead, the type-checker assumed non-empty tuple types, resulting in misleading behavior. The fix added an explicit check in the type computation logic to detect empty tuple types and return an informative error message: \texttt{"Join requires non-nullary relations"}. This patch aligns the implementation with relational semantics and improves diagnostics for users. The test demonstrates how corner cases in relational encodings, like nullary joins, can reveal overlooked semantic constraints.

Figure~\ref{sample:c} presents a bug in Z3's model evaluation logic, where dereferencing a null pointer led to a crash when evaluating partially interpreted functions without default expressions. The crafted input includes divisions by zero and function applications that trigger the creation of partial function interpretations with entries but no associated body. The evaluator assumed the presence of an interpretation term and attempted substitution without checking for null, resulting in a segmentation fault. The patch introduced a guard that skips substitution if the interpretation is null, safely falling back to alternative handling. This case illustrates how corner-case arithmetic and partial models can interact to produce undefined internal states, emphasizing the importance of defensive programming in model evaluation.

\section{Discussion}
\label{sec:discussion}

\subsection{Impact of  the Self-Correction Mechanism}

\tool~incorporates a self-correction mechanism to ensure the validity of generated formulas. This mechanism serves several crucial purposes: it enhances the quality of our formula generators, mitigates the \textit{hallucination} tendency of LLMs which can produce nonsensical outputs, and ultimately boosts the overall effectiveness of \tool. 
To assess the impact of the correction mechanism, we measured the proportion of syntactically valid formulas generated before and after its application.
Initially, generators derived directly from summarized context-free grammars often produced a high number of invalid formulas that could not be successfully parsed by either cvc5 or Z3.
For instance, in syntactically intricate theories like finite fields and newly introduced higher-order theories, the proportion of valid formulas was often below 30\%. In contrast, theories like real arithmetic fared much better, with over 90\% valid formulas.
Our correction mechanism substantially improved these outcomes. 
For most theories, the proportion of valid formulas increased to nearly 100\%. 
Even in the more challenging and recently added theories, validity improved markedly to over 80\%. These results demonstrate that the self-correction mechanism plays a critical role in increasing formula validity, reducing the number of malformed inputs that could otherwise impede bug discovery.
Furthermore, \tool~significantly outperforms the direct use of LLMs for test input generation. As shown in prior work~\cite{DBLP:journals/corr/abs-2308-04748}, relying solely on LLMs can lead to over 50\% of formulas being syntactically invalid. In contrast, our approach, with the correction mechanism in place, yields far more useful test inputs.

    \subsection{Threats to Validity}
    This study is susceptible to several primary threats that warrant consideration. 
    
    \mypara{Internal.}
    The main internal threat is that the presence of randomness in the fuzzing process could introduce variability in the results of our experiments. 
    To minimize the impact of non-deterministic behavior, we followed established experimental protocols from prior literature~\cite{sunsmt, 10.1145/3622781.3674171}. Specifically, we employed  long execution windows for code coverage collection to ensure that our results represent stable trends.
    Another concern is that our evaluation primarily focused on default solver configurations. Because many internal code paths are disabled or inaccessible via the command-line interface under default settings, our results may not reflect the absolute maximum coverage each fuzzer could achieve. However, because our main objective is to perform a fair, head-to-head comparison between tools rather than to maximize absolute coverage, this setup  provides a consistent baseline for our observations.

\mypara{External.}
A primary external threat is the potential for the LLM to generate inaccurate outputs, commonly referred to as ``hallucinations'', when summarizing context-free grammars or implementing generators. Such inaccuracies may result in invalid or ineffective test formulas. To address this issue, we integrate a self-correction mechanism into \tool, which enhances the validity of the generated formulas.

\subsection{Limitation and Future Work}

\tool~has several limitations that suggest directions for future work.
First, \tool~requires synthesized generators to produce Boolean terms. This simplifies mutation and ensures well-typed SMT-LIB formulas, but may limit the diversity of generated inputs. Extending \tool~to support other term types (e.g., bit-vectors, reals, arrays) or mixed-sorts expressions could broaden the search space and uncover additional classes of solver bugs.
Second, the current self-correction mechanism emphasizes syntactic validity. It improves conformance to inferred grammars but does not explicitly promote semantic diversity or bug-finding effectiveness. Incorporating solver-driven signals, such as coverage feedback, exploration of rare configurations, or divergent solver behaviors, could guide generators toward more meaningful and stress-inducing formulas.
Third, \tool~relies on solver documentation to summarize grammars, but such documentation is often incomplete or informal. This can introduce inaccuracies that reduce generator quality. Future work could draw on alternative sources, including solver codebases, regression tests, SMT-LIB benchmarks, or grammar information inferred directly from solver behavior.
Finally, our evaluation focuses on SMT solvers and several advanced LLMs. As new models emerge, it will be valuable to study how model choice and prompting strategies affect validity, diversity, and bug discovery. More broadly, the methodology behind \tool~may extend to other domains that use grammar-based or template-driven testing, such as compilers and interpreters. 
In addition, we plan to further automate the current semi-automatic bug triage process. In particular, we aim to investigate automated identification of bug-inducing commits and explore how LLMs can assist in diagnosing recurring or similar issues.

\section{Related Work}
\label{sec:related-work}

\mypara{SMT Solver Fuzzing.}
SMT solvers are widely adopted in modern systems for verification, testing, and synthesis across the systems and programming languages communities~\cite{DBLP:conf/sosp/JiaPTWZA19,DBLP:conf/sosp/NelsonSZJBTW17,DBLP:conf/osdi/HawblitzelHLNPZZ14,DBLP:conf/osdi/HanceLHHJP20,DBLP:conf/asplos/HanWQSXWZYL24,DBLP:conf/nsdi/LopesBGJV15,DBLP:conf/nsdi/SinghaMBKMV24}.
Given their widespread adoption and critical role, ensuring the correctness and robustness of SMT solvers is essential, which has motivated extensive research on solver testing and fuzzing.
To improve their reliability, grammar-based fuzzers such as ET~\cite{DBLP:journals/pacmpl/Winterer024} and Murxla~\cite{DBLP:conf/cav/NiemetzPB22} were introduced, but both rely highly on expert-crafted generation rules. 
Mutation-based approaches~\cite{winterer2020validating,mansur2020detecting,DBLP:journals/pacmpl/WintererZS20,DBLP:journals/pacmpl/ParkWZS21} provide an alternative.
For instance, OpFuzz~\cite{DBLP:journals/pacmpl/WintererZS20} mutates the operators in seeds to generate new test formulas.
HistFuzz~\cite{sunvalidating} proposes to exploit skeletons of historical bug formulas to produce mutants.
However, these methods remain constrained by manual mutation strategies. 
Our work instead leverages LLMs to synthesize reusable term generators across different theories, including newly updated ones. 
These generators aim to fill skeletons with diverse terms to produce effective test cases. 
Our tool, \tool, has demonstrated effectiveness on state-of-the-art solvers across both standard and solver-specific theories.

\mypara{Large Language Models.}
LLMs have achieved remarkable success in generation and reasoning~\cite{DBLP:journals/corr/abs-2307-09288}, and have been applied to software engineering for code generation, completion, and testing~\cite{DBLP:conf/icse/WangJLYX0L22,sunsmt,DBLP:journals/corr/abs-2308-04748,DBLP:conf/pldi/YeTTHFSBW021,deng2023large,DBLP:journals/corr/abs-2304-02014}. 
For example, Ou et al.~\cite{10.1145/3622781.3674171} fuzz compilers with LLM-designed mutators, and Wang et al.~\cite{DBLP:journals/corr/abs-2507-19275} further investigate this paradigm.
LLMs have also been applied to SMT solver fuzzing~\cite{sunsmt,DBLP:journals/corr/abs-2308-04748}, such as LaST~\cite{sunsmt}, which retrains LLMs for formula generation. 
However, direct generation often yields invalid inputs and requires costly retraining or reasoning.
Concurrent work, G$^2$FUZZ~\cite{DBLP:conf/uss/ZhangLW0025}, also integrates LLMs into fuzzing but differs substantially from our approach. 
It prompts LLMs to produce deterministic scripts that generate a small set of fixed inputs and relies on AFL++ for mutation. 
Such mutation is not well suited to the highly structured nature of SMT formulas.
Moreover, because each generator produces only a handful of inputs, G$^2$FUZZ repeatedly queries the LLM to obtain new generators, incurring recurring latency and nontrivial cost.
In contrast, \tool queries the LLM once per theory to synthesize reusable random generators grounded in solver documentation. 
These generators systematically complete skeletons, avoid repeated LLM interaction, and reduce invalid cases, thereby improving the overall efficiency and scalability of SMT solver fuzzing.

\mypara{Skeleton-based Testing.}
Skeleton- or template-based testing is effective in exposing deep bugs. 
For example, Skeletal Program Enumeration (SPE)~\cite{DBLP:conf/pldi/ZhangSS17,10.1145/3647994} tests compilers via systematic skeleton enumeration, while later work applied templates to MLIR compilers~\cite{DBLP:conf/kbse/WangCXLWSZ23} and JVMs~\cite{DBLP:conf/sosp/Li00S23}. 
These techniques, however, require manually designed strategies or skeletons. 
Our method reduces this effort by automatically completing skeletons with LLM-synthesized generators. 
This strategy has already uncovered \confirmed~confirmed bugs in solvers and shows promise for adoption in other domains.

\section{Conclusion}
\label{sec:conclusion}

This paper introduces \tool, an innovative approach for testing SMT solvers that leverages the capabilities of LLMs.
Instead of directly using LLMs, we employ them to design and implement generators for various SMT theories, including newly added and solver-specific ones.
Furthermore, \tool~synthesizes the terms generated by LLM-produced generators into skeletons extracted from real-world formulas.
We implemented \tool~as a practical fuzzing tool and conducted a comprehensive evaluation using two well-known SMT solvers, namely Z3 and cvc5. 
Notably, \tool~has identified \confirmed~confirmed real bugs in the solvers, with \fixed~of them already fixed by developers. 
Moreover, our experiments demonstrate that \tool~outperforms state-of-the-art SMT solver fuzzers in terms of code coverage and bug-finding ability.
This study demonstrates an effective approach to utilizing LLMs for software system testing, particularly for rapidly evolving systems.

\section*{Acknowledgments}
We are grateful to the anonymous reviewers for their valuable and constructive feedback.
We also express our sincere gratitude to our shepherd, Siva Kesava Reddy Kakarla, for his insightful guidance and support during the revision of this work.
We also greatly appreciate the Z3 and cvc5 development teams, in particular Nikolaj Bj{\o}rner, Andrew Reynolds, and Clark Barrett, for their support and for carefully examining the bugs we reported.
This work was supported in part by the National Natural Science Foundation of China (Grants 624B2067, 62572226, and 62472215), the Jiangsu Natural Science Foundation (Grant BK20231402), the Collaborative Innovation Center of Novel Software Technology and Industrialization.

\normalem
\bibliographystyle{ACM-Reference-Format}
\balance
\bibliography{full}

\end{document}